\documentclass[useAMS,usenatbib]{mnras}

\usepackage{graphicx}	
\usepackage{amsmath}	
\usepackage{amssymb}	
\usepackage{multicol}        
\usepackage{bm}		

\newcommand{\dif}{\mathrm{d}}
\newcommand{\Pc}{P_{\mathrm{c}}}

\newcommand{\Myr}{\mathrm{Myr}}
\newcommand{\Gyr}{\mathrm{Gyr}}
\newcommand{\AU}{\mathrm{au}}
\newcommand{\MJ}{{\mathrm{M_J}}}
\newcommand{\MSol}{\mathrm{M_{\sun}}}
\newcommand{\RJ}{\mathrm{R_{J}}}
\newcommand{\RSol}{\mathrm{R_{\sun}}}
\newcommand{\sbp}{\bar{\sigma}_{\mathrm{pl}}}
\newcommand{\sbs}{\bar{\sigma}_{\star}}

\title[The Eccentricities of Youth]{Constraining Planetary Migration and Tidal Dissipation with Coeval Hot Jupiters}

\author[O'Connor \& Hansen]
{Christopher E. O'Connor$^{1}$\thanks{E-mail: ceoconnor@g.ucla.edu} and Bradley M. S. Hansen$^{1,2}$ \\
$^{1}$Department of Physics and Astronomy, University of California, Los Angeles, Los Angeles, CA 90095, USA \\
$^{2}$Mani L. Bhaumik Institute for Theoretical Physics, University of California, Los Angeles, Los Angeles, CA 90095, USA}

\begin{document}

\date{Submitted on 27 November 2017.}


\maketitle

\label{firstpage}

\begin{abstract}
    We investigate the constraints on the formation of, and tidal dissipation processes in, hot Jupiters (HJs) that can be inferred based on reliable knowledge of the age of a system or population. Particular attention is paid to the role of young systems (such as those in open clusters or star-forming regions) in such studies. For an ensemble of coeval HJ (or proto-HJ) systems, we quantify the effect of age on the distribution of orbital eccentricities with respect to orbital periods as well as the location of the observed ``pile-up'' feature. We expect the effects of pre-main-sequence stellar evolution to be important only if a substantial fraction of HJs approach their current orbits early in protostellar contraction (ages $\la 10 \, \Myr$). Application to the HJs presently known in the cluster M\,67 yields constraints on the dissipation roughly consistent with those gleaned from planets in the field; for those in the Hyades and Praesepe, our results suggest a higher degree of dissipation at early times than that inferred from other populations.
\end{abstract}


\section{Introduction}

The dynamical histories of many extrasolar planetary systems seem inevitably to include tidal interactions between planets and stars. More than two decades after the discovery of 51\,Peg\,b \citep{mq95}, circular orbits clearly predominate among planets with short orbital periods while a progressively broader range of eccentricities is observed at longer periods. Despite the topic's lengthy history \citep[see review by][]{ogilvie14}, both the overall strength of tidal effects and the specific mechanisms of dissipation are still a subject of active debate.

Tidal effects operate on long timescales, which makes the study of tidal evolution sensitive to the ages of the systems under consideration. Most exoplanets have so far been discovered around main-sequence stars in the field, which can be assigned an approximate age based on stellar isochrones and evolutionary models; but these age estimates typically feature broad error bars. Relatively few planets have been discovered around stars younger than $1 \, \Gyr$ or so. These facts jointly have made it difficult to distinguish between various formation channels of specific planetary architectures---for instance, between the disc-driven migration and high-eccentricity migration paradigms for the formation of hot Jupiters (HJs; we define these planets to be more than a third the mass of Jupiter and to orbit in less than ten days). Because disc-driven migration occurs on time-scales much shorter than does high-eccentricity migration ($\la 10 \, \Myr$ versus $\ga 100 \, \Myr$), one would desire a robust sample of young HJ systems in order to better constrain which of these two pipelines may be dominant.

The discovery of HJs residing in coeval stellar populations, especially open clusters \citep{quinn+12,quinn+14,brucalassi+14,brucalassi+16,rizzuto+17}, may be able to break this degeneracy on both fronts: clusters have more precise age estimates than stars in the field; and, since most clusters dissolve within $\sim 1 \, \Gyr$ of formation, they tend to lie in the very age range during which high-eccentricity migration might be observed. While relatively few giant planets have yet been discovered in clusters (see \citealt*{quinn+15}; \citealt{quinn16}; and references therein), it is worth considering the utility of such objects as the sample inevitably grows.

In this paper, we seek to quantify the extent to which the discovery of cluster-member HJs and other coeval HJ populations can help to constrain the strength of tidal interactions in these systems. This, in turn, can help to shed light on the mechanisms that drive planetary migration. In Section \ref{s:tides}, we review our preferred formalism for quantitative discussions of tidal dissipation. In Section \ref{s:PEA}, we discuss the features of a coeval planet population which may be useful for constraining the dissipation in those planets. In Section \ref{s:disc}, we discuss constraints from the existing sample of coeval and young HJs and prospects for the improvement of this sample with current and future exoplanet surveys. We review our primary results in Section \ref{s:conc}.

\section{Tidal Formalism} \label{s:tides}

Given the complexity of tidal interactions \citep{ogilvie14} and the further complications introduced by stellar and planetary evolution, it is difficult to treat the evolution of HJ orbits under tidal dissipation in a fully consistent manner. The most widespread approximation for the problem is that of the equilibrium tide \citep*{hut81,ekh98}. We adopt the incarnation of this model derived by \citet{ekh98}, as revised and calibrated by \citeauthor{h10} (\citeyear{h10}, \citeyear{h12}; hereafter H10 and H12, respectively). In this framework, the dissipation strength in a tidally distorted body is expressed in terms of a normalized bulk dissipation constant, $\bar{\sigma}$. This is defined by
\begin{equation} \label{eq:sigmaDef}
    \sigma \equiv \bar{\sigma} \left( \frac{G}{M R^{7}} \right)^{1/2},
\end{equation}
where $M$ ($R$) is the mass (radius) of the distorted body and $G$ the gravitational constant. For stars, $\sbs$ is calculated relative to the mass and radius of the Sun; the constant $\sbp$ for planets is calculated similarly relative to Jupiter.

The bulk dissipation constant proper, $\sigma$, is related to the characteristic time-scale of orbital evolution by
\begin{equation} \label{eq:tauDef}
    t_{\rm F} = \frac{1}{9 \sigma} \frac{M}{m (M + m)} \frac{a^{8}}{R^{10}},
\end{equation}
where $m$ is the mass of the perturbing companion and $a$ is the semimajor axis of the osculating orbit. We find $\sigma$ to be more convenient for our task than more traditional measures of tidal dissipation, namely the quality factors $Q$ and $Q'$, because $\sigma$ has a weaker dependence on ``extrinsic'' properties of the planet--star system (especially the forcing frequency) than do the quality factors.

In their derivation, \citet{ekh98} provide for tidal interactions between misaligned orbits and spins, and indeed many HJ systems in the field exhibit spin--orbit misalignments \citep{winn+10,albrecht+12}. However, while the evolution of the stellar spin is governed primarily by dissipation in the star, circularisation of the planetary orbit is expected to be dominated by dissipation in the planet; the latter is largely independent of the orientation of the orbit. Therefore, although we include the effects of dissipation in the star in our computational models, we consider only aligned systems for the sake of simplicity. The equations of tidal evolution are given, in this limit, by equations (1), (3), and (4) of H10. In this section, we do not consider either stellar or planetary evolution; for a discussion of these additional effects, we refer the reader to Section \ref{s:PEA:SSE}. In any case, stellar evolution ought to be a relatively minor component during the first billion years or so after the zero-age main sequence (the typical lifetime of a galactic open cluster) for FGK stars.

\subsection{Revised Dissipation Values for the Field} \label{s:tides:calib}

In order to establish a baseline of comparison for the effects of age, we present, in Appendix \ref{s:field}, an update to the calibration of our model using the population of giant planets around solar type stars in the field. We calibrate the expected degree of dissipation in both planet and star by applying the procedures followed by H10 to the sample of transiting giant planets studied by Bonomo et al. (\citeyear{bonomo+17}; hereafter B17). In brief, we constrain the rate of dissipation by requiring the eccentricities of observed HJs to be consistent with a model population at the estimated ages of the systems in question. The resultant estimate for the average normalized dissipation constant for giant planets is
$$ \sbp = (8 \pm 2) \times 10^{-7}, $$
with the caveat that one would expect real planets to show variance from this value by perhaps an order of magnitude (e.g., H12). The stellar dissipation constant (which we estimate using the most-massive planets in the sample) displays more variance between systems than can be accommodated by a single, mean value (see Sections \ref{s:field:indiv:WASP18} and \ref{s:field:indiv:WASP89}). Since we are interested more in the planetary dissipation, and since stellar tides begin to be quantitatively important only for planets greater than a few Jupiters in mass (H10), we adopt a fiducial stellar dissipation constant of $\sbs = 10^{-7}$, which straddles several constraints on this value by H10, H12, and \citet{penev+16}.

\section{Tides in Time} \label{s:PEA}

\begin{figure}
    \includegraphics[width=\columnwidth]{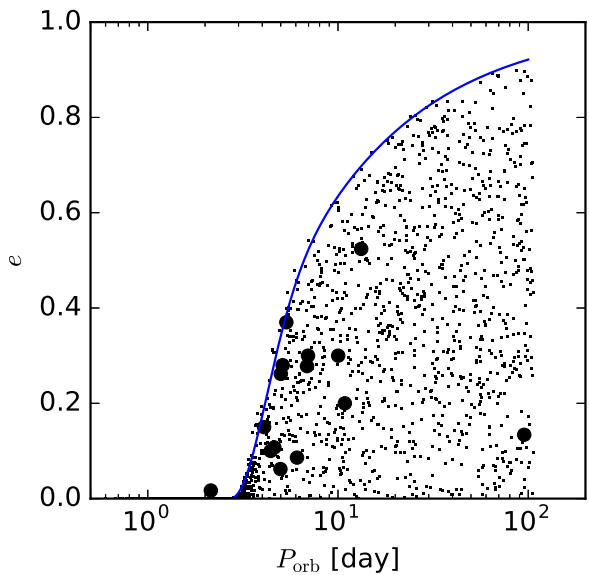}
    \caption{A typical $P$--$e$ diagram. The large dots are various giant planets from the B17 and open-cluster samples. The small points are a model population of planets drawn from a uniform distribution in $\log a$ and $e$, so as to best illustrate the region of parameter space where eccentric planets may be found. This particular population was evolved to an age of $3 \, \Gyr$ with $\sbp = 3.0 \times 10^{-6}$. The blue contour is an approximate fit of equation (\ref{eq:PE_env}) to this population, with $\Pc = 2.7 \, \dif$, $P_{\rm ecc} = 2.2 \, \dif$, and $\beta = 1.0 \, \dif^{-1}$.}
    \label{fig:PE_ex}
\end{figure}

Figure \ref{fig:PE_ex} depicts the typical appearance of a coeval HJ population in the $P$--$e$ plane (with the $\sbp$ value assumed above). For comparison to observations, we plot various planets in the B17 and cluster samples, which shows basically the same trend of increasingly large eccentricities with period. The distribution of eccentricities with respect to orbital periods (hereafter the ``$P$--$e$ relation'' or ``$P$--$e$ distribution'') has been studied previously with tidal circularisation in mind. Indeed, \citeauthor{mm05} (\citeyear{mm05}; MM05) studied the circularisation of close binary stars in coeval stellar populations, a task similar in spirit to that which we take up for coeval HJs. They found that the {\em average eccentricity} of a coeval binary population at a given orbital period was described well by a function of the form
\begin{equation} \label{eq:MM05}
    \bar{e}(P) =
        \left\{
        \begin{array}{ll}
            \alpha [1 - \exp(\beta (\Pc - P))], & P \geq \Pc; \\
            0, & P < \Pc.
        \end{array}
        \right.
\end{equation}
As MM05 intended it, the so-called ``circularisation period'' ($\Pc$) is a measure of the average degree of circularisation in a coeval population. For exoplanets, it often corresponds closely to the period of the smallest orbit for which nonzero eccentricities persist at the age of the population. MM05 showed that $\Pc$ varies roughly as a function of age for solar-type binaries in clusters; presumably, a similar effect would be seen among eccentric HJs and proto-HJs in clusters if enough of them were discovered. H10 found a value of $\Pc \approx 3 \, \dif$ for giant planets orbiting stars in the field; as most of these systems are well over $1 \, \Gyr$ old, this sets an approximate upper limit on the degree of circularisation we would expect among HJs in a typical star cluster.

\begin{figure}
    \includegraphics[width=\columnwidth]{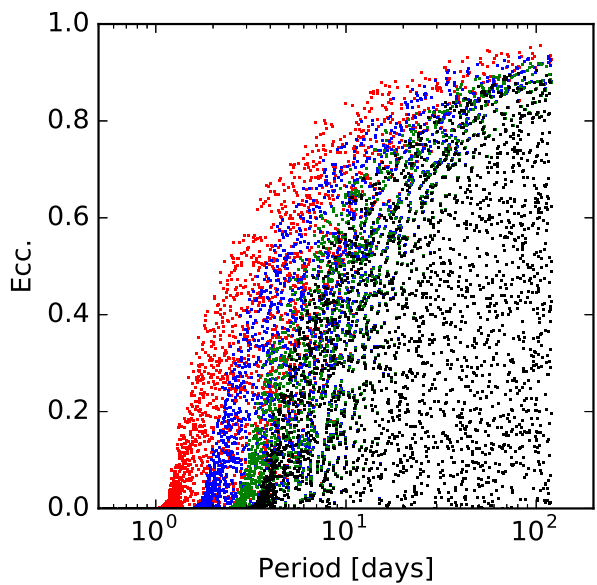}
    \caption{The evolution of the $P$--$e$ distribution with age. A single model population of $1 \MJ$ planets is show at ages of $30 \, \Myr$ (red), $300 \, \Myr$ (blue), $3 \, \Gyr$ (green), and $11 \, \Gyr$ (black). When the horizontal axis has a logarithmic scale, the appearance of the distribution remains invariant with age; it merely moves rightwards as orbits with longer periods circularise. Note the pile-up of planets near the circularisation period.}
    \label{fig:PE_evo}
\end{figure}

For the purposes of exoplanet studies, MM05's $P$--$e$ distribution has the shortcoming that, within a given cluster, the number of HJ systems is unlikely ever to be comparable to the number of binary stars. Although \citet{brucalassi+16} found some evidence for an enhanced HJ fraction in the solar-age cluster M\,67 (see Section \ref{s:disc:ex:M67} below), their estimate is still rather smaller than the binary fraction of that cluster \citep*[e.g.,][]{mathieu+90}, and it is unclear as yet whether this finding was coincidental. It may be more useful for our purposes to examine the nature of the upper envelope of the $P$--$e$ distribution---that is, the maximum eccentricity that could be observed at a given orbital period and age. We have tested various functions as approximations to this shape, and find the best candidate to be:
\begin{equation} \label{eq:PE_env}
    e_{\rm max} (P) = \left[ 1 - \left( \frac{P}{P_{\rm ecc}} \right)^{-2/3} \right] \left[ 1 - \exp\left( -\beta \left( P - \Pc \right) \right) \right],
\end{equation}
for $P \geq \Pc$ and $e_{\rm max}(P) = 0$ for $P < \Pc$. For a visual comparison of this equation to the actual envelope of a model population, see Figure \ref{fig:PE_ex}

In the case that stellar tides are negligible, one would expect the evolution of a single planet in $P$--$e$ phase space to track a contour of constant angular momentum, $a(1 - e^{2}) = {\rm const}$. However, because we are interested more in the evolution of an ensemble of planets, of greater importance to us is the threshold for significant tidal evolution over time-scales of a $\Gyr$ or more. Therefore, in equation (\ref{eq:PE_env}) the first bracketed term corresponds to a contour of constant periastron distance, $a(1 - e) = {\rm const}$. This is motivated by the conventional wisdom that a planet must pass within a characteristic distance $\la 0.05 \, \AU$ of its star in order to circularize within a few $\Gyr$; this term dominates the $P$--$e$ envelope for eccentricities greater than $0.5$ or so. The second bracketed term is inspired by equation (\ref{eq:MM05}). Its purpose is to capture the behavior of the envelope at low eccentricities, where it tapers to zero at a period somewhat longer than that expected from extrapolating the constant-periastron behavior; it is for that reason that two characteristic periods, $\Pc$ and $P_{\rm ecc}$, are introduced. In the right panel of Figure \ref{fig:PE_ex}, we compare the shape of equation (\ref{eq:PE_env}) to another example population.

To investigate the evolution of the $P$--$e$ distribution with time, we have simulated the orbital evolution of a large sample ($N = 4000$) of identical giant planets due to equilibrium tides raised on both planets and host stars. The initial conditions of the model planets were chosen from uniform distributions in $\log a$ and $e$, so as to probe the full range of possible initial conditions. Each starting point was integrated forward using equations (1), (3), and (4) of H10 for the specified age of the population. In Figure \ref{fig:PE_evo}, we depict the population at four ages: $30 \, \Myr$ and $300 \, \Myr$, typical of galactic open clusters and moving groups; $3 \, \Gyr$, typical of Population I stars in the field; and $11 \, \Gyr$, the ages of globular clusters and the oldest field stars.

Upon inspection of Figure \ref{fig:PE_evo}, it is immediately apparent that the younger populations display a broader range of eccentricities at shorter orbital periods than do the older. This can scarcely be considered a surprise, but it quantitatively validates the assertion that young HJs potentially can provide meaningful evidence for high-eccentricity migration, as well as constraints on tidal dissipation.

\subsection{The Period Pile-up} \label{s:PEA:pileup}

\begin{figure*}
    \includegraphics[width=\columnwidth]{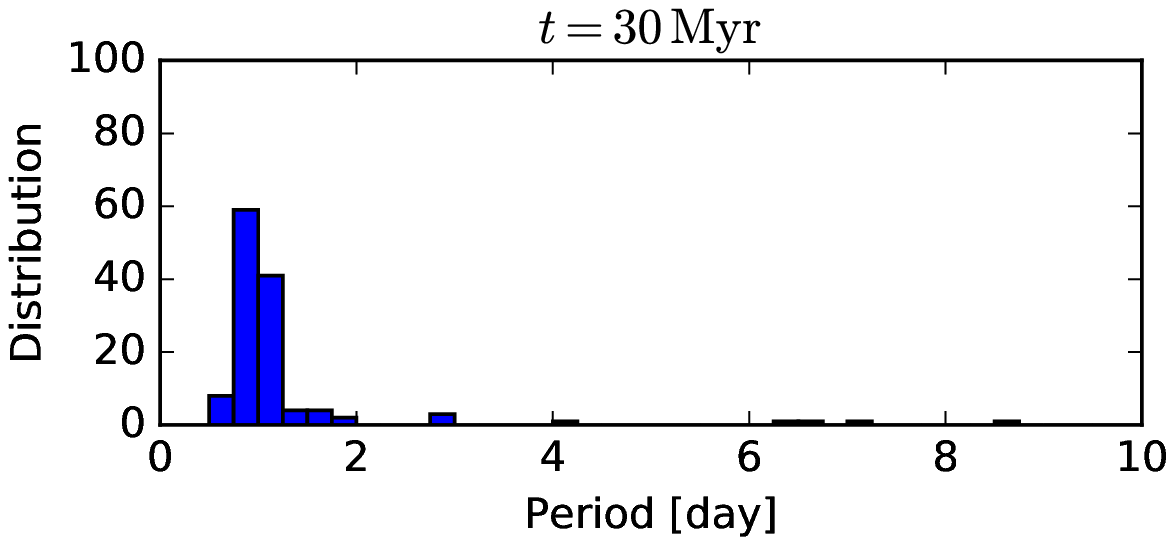}
    \includegraphics[width=\columnwidth]{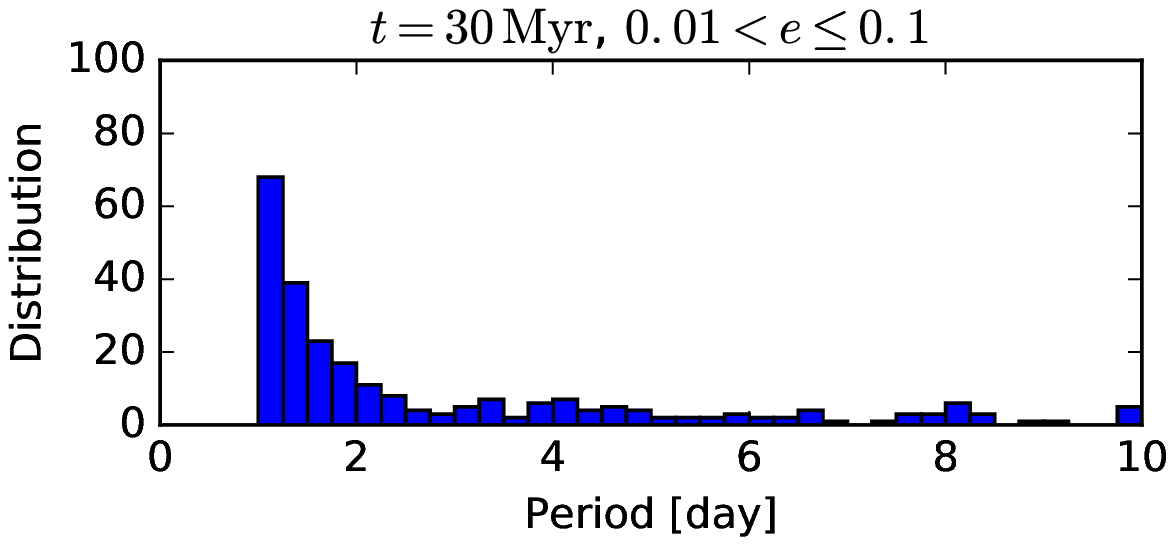}
    \includegraphics[width=\columnwidth]{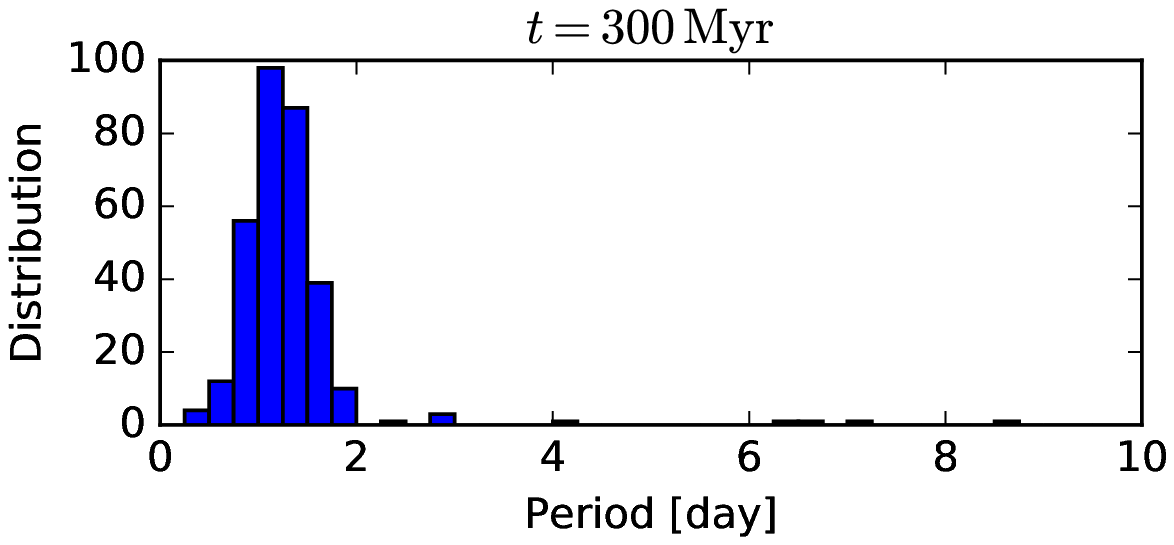}
    \includegraphics[width=\columnwidth]{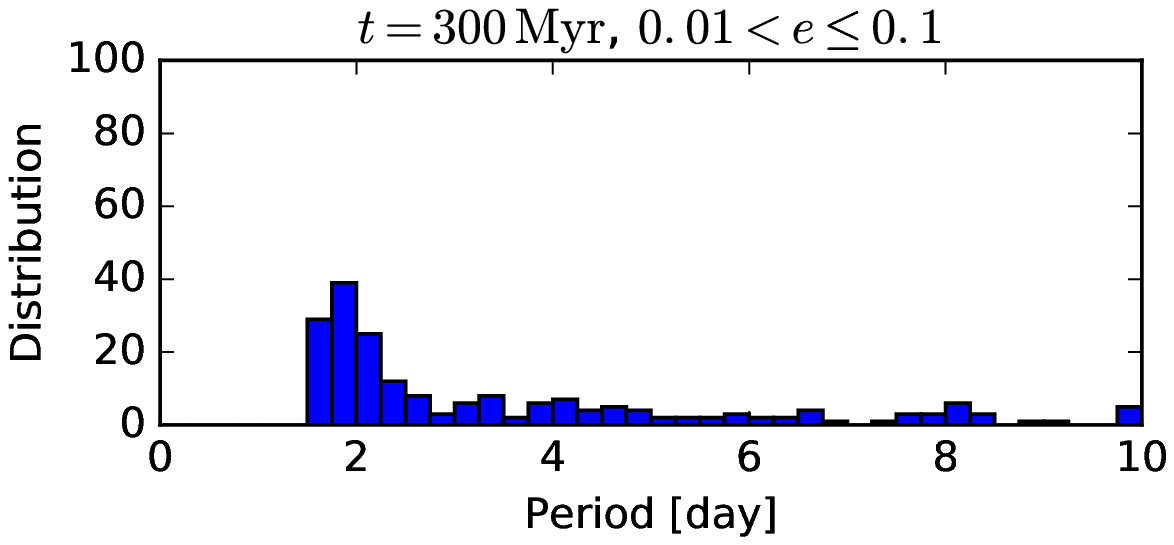}
    \includegraphics[width=\columnwidth]{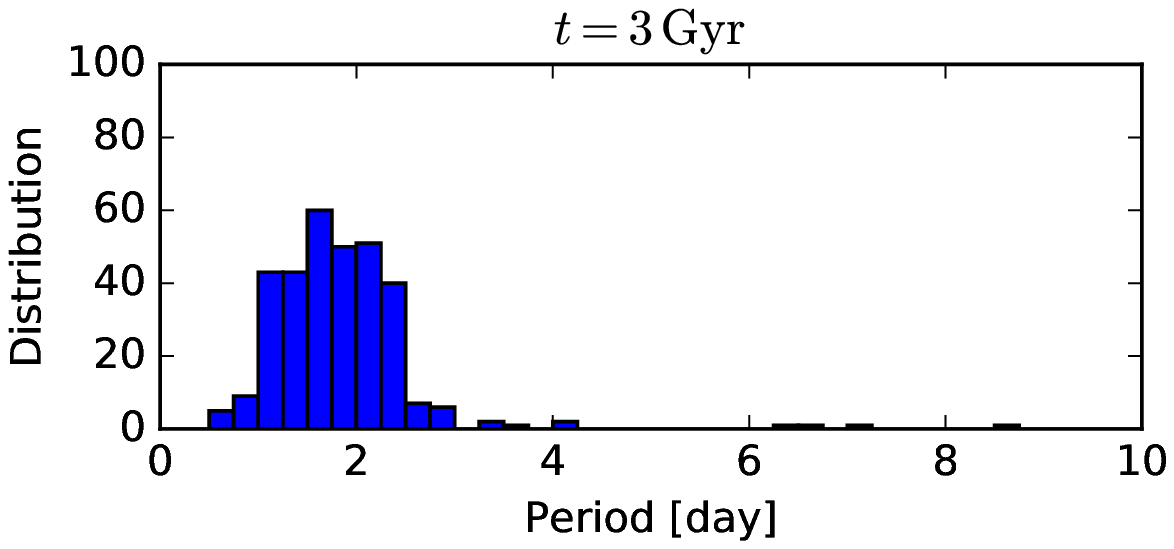}
    \includegraphics[width=\columnwidth]{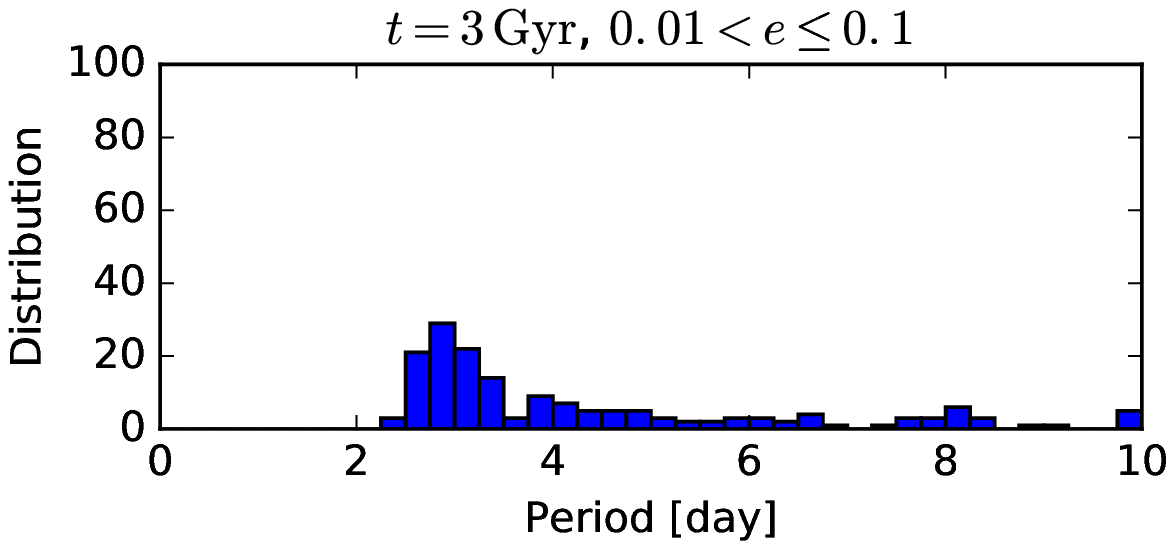}
    \includegraphics[width=\columnwidth]{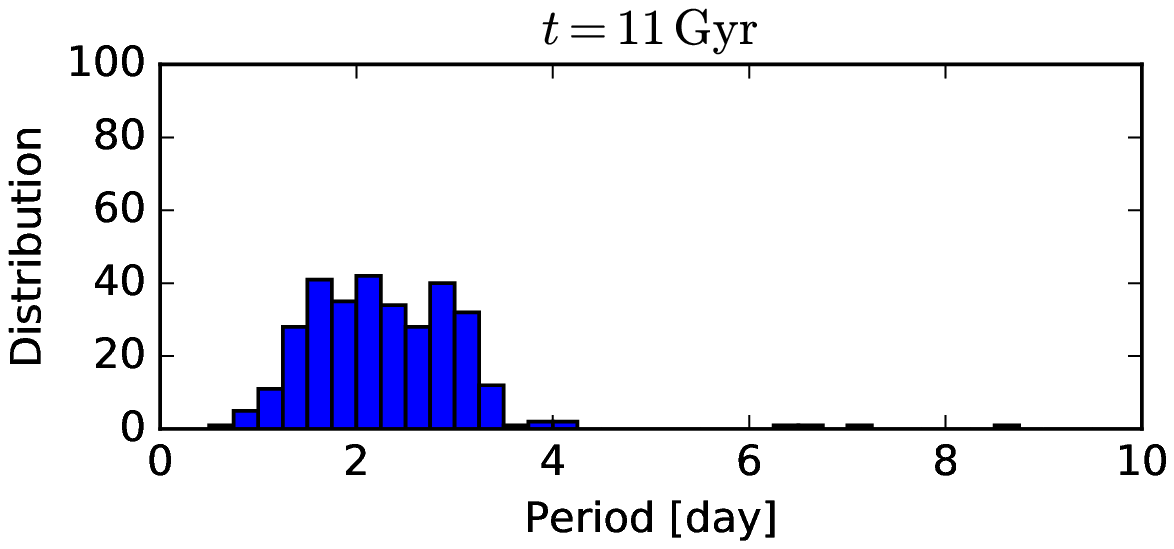}
    \includegraphics[width=\columnwidth]{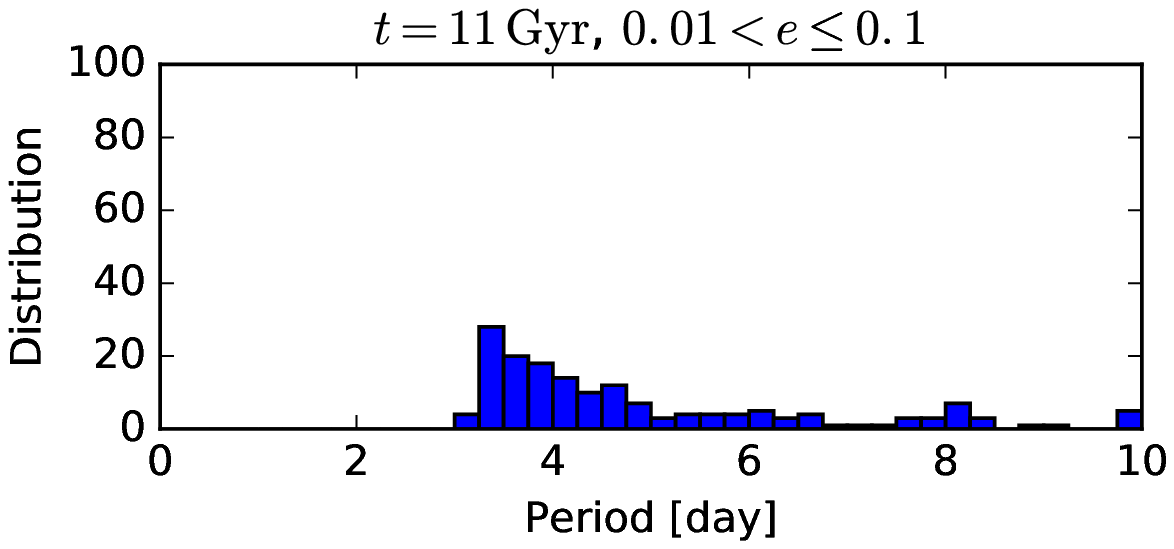}
    \caption{Histograms of the orbital periods of our model planets with $e \leq 0.01$ (left column) and $0.01 < e \leq 0.1$ (right), at ages of $30 \, \Myr$ (top row), $300 \, \Myr$ (upper middle), $3 \, \Gyr$ (lower middle), and $11 \, \Gyr$ (bottom).}
    \label{fig:PHist}
\end{figure*}

One of the most notable features of Figure \ref{fig:PE_evo} is that our model population naturally forms an over-density of planets at periods just above the circularisation period. This is a consequence of the fact that equilibrium-tidal evolution maps a wide range of initial $a$ and $e$ into a single value of final semi-major axis, depending on the total angular momentum. Just such a feature is observed in the period distribution of observed HJs around $P \sim 3 \, \dif$ \citep*{gaudi+05}.

In Figure \ref{fig:PHist}, we show histograms of our model planets with respect to orbital period, for two ranges of planetary eccentricities: planets which have circularised to eccentricities $e \leq 0.01$, and planets in the range $0.01 < e \leq 0.1$ which are still in the process of circularisation. Importantly, we exclude planets that orbit within their Roche limits,
\begin{equation}
    a_{\rm Ro} \approx 8.06 \times 10^{-3} \, \AU \left( \frac{R_{\rm pl}}{\RJ} \right) \left( \frac{\mu}{10^{-3}} \right)^{-1/3},
\end{equation}
where $\mu \equiv M_{\rm pl} / M_{\star}$; this may occur as the semimajor axis slowly shrinks due to the action of stellar tides.

The most notable feature of Fig. \ref{fig:PHist} is the clear presence of a pile-up at every age, in both eccentricity ranges. The evolution of the pile-up of circular orbits reflects the competition between planetary tides---which act to circularise the orbit and extend the outer edge of the pile-up to longer orbits---and stellar tides---which act to drag planets inwards, slowly removing them at the inner edge. The pile-up for orbits with moderate eccentricities reflects only circularisation due to planetary tides---its inner edge coincides roughly with $\Pc$ at each age. In tandem, these two, related structures might be used as quasi-independent constraints of $\sbp$ and $\sbs$.

Interestingly, the piles-up among the circularised populations at $30 \, \Myr$ and $300 \, \Myr$ are located at significantly shorter orbital periods than that observed in the field today: at $30 \, \Myr$, it is localised around $P \approx 1 \, \dif$; at $300 \, \Myr$, it extends to $\approx 2 \, \dif$. The precise value is a function of both our choice of $\sbp$ and our initial population model, but the qualitative dependence of the pile-up on age is robust and can potentially provide useful insights on tidal evolution.

Finally, we note that the diminution of the HJ population at short orbital periods depends on the evolution of the stellar spin with age; this is because a planet can only fall into its host star if that star spins more slowly than the planet orbits. However, it is understood that young, cool stars spin much more rapidly than their mature counterparts, due to the effects of magnetic braking (\citealt{barnes10} and references therein). According to \citeauthor{barnes10}' rotational isochrones, typical FGK stars can be expected to spin with periods shorter than $\sim 2$ days until an age of at least $\sim 100 \, \Myr$; in that case, the youngest, shortest-period HJs might move {\it outward} under tidal torques, which also could contribute to a pile-up. However, by $\sim 600 \, \Myr$, most of these stars would have slowed their spins to periods greater than $10$ days. Thus, their HJs would have begun to migrate inwards by that time, and the outcome at ages of several $\Gyr$ would be largely the same as before.

\subsection{The Circularisation--Age Relation} \label{s:PEA:age}

\begin{figure}
    \includegraphics[width=\columnwidth]{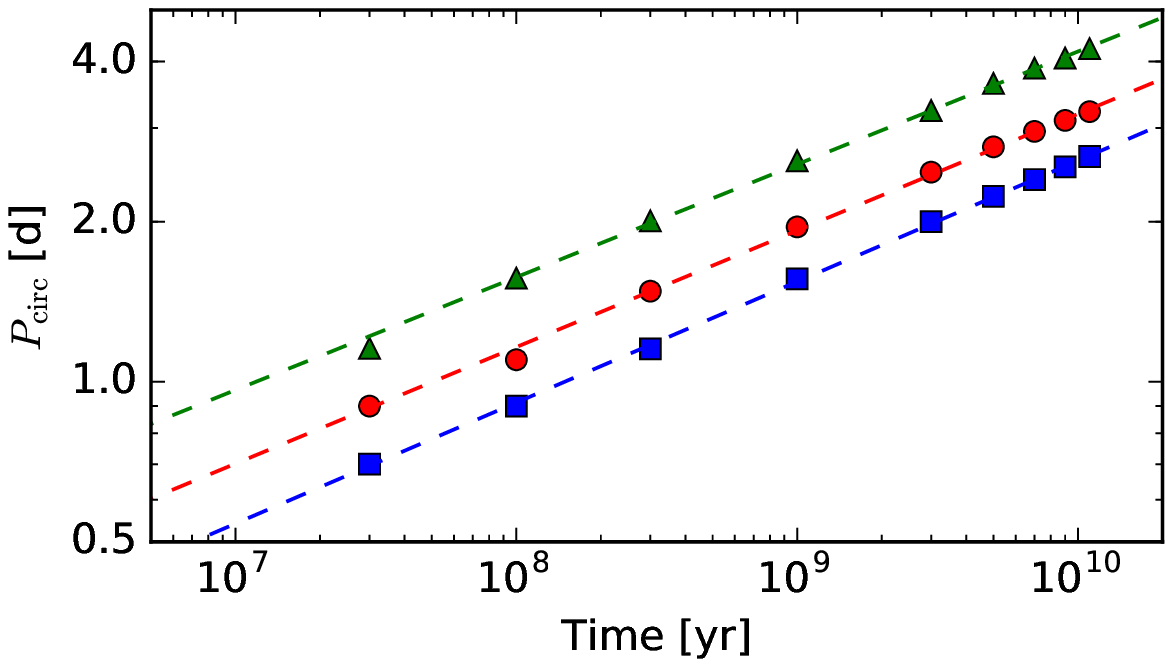}
    \caption{The circularisation period as a function of age for our $1 \MJ$ model planets, using three different planetary dissipation constants: our ``in-house'' calibration of $\sbp = 8 \times 10^{-7}$ (red dots), H10's result of $\sbp = 3 \times 10^{-7}$ (blue squares), and the value $\sbp = 3 \times 10^{-6}$ we infer later for the HJs of M\,67 (green triangles). The best-fitting pure power law (equation \ref{eq:PcTAff} with $P_{0} = 0$) is displayed as a dashed line of the same color in each case. Notice that the fits are somewhat looser at ages less than $\sim 100 \, \Myr$.}
    \label{fig:PcAge}
\end{figure}

Notice that in Figure \ref{fig:PE_evo}, the first three distributions (red, blue, and green from left to right) are spaced by a factor of ten in age from their neighbors. When viewed in a logarithmic scale on the period axis, as we have done, these distributions move by a uniform amount to the right---that is, their circularisation period increases with by a constant multiplicative factor. This suggests that, to good approximation, the circularisation period obeys a power law with age:
$$ \Pc(t) \propto t^{\gamma}. $$
In principle, the size of an orbit is bounded from below by the Roche limit of a planet, and so one might allow for a ``zero-age'' circularisation period, $P_{0}$, appearing as a constant offset:
\begin{equation} \label{eq:PcTAff}
    \Pc(t) = k_{P} \left( \frac{t}{1 \, \Gyr} \right)^{\gamma} + P_{0}.
\end{equation}
However, we do not find that fits allowing for nonzero $P_{0}$ are substantially better; henceforth, we will set $P_{0} = 0$.

Different values of the dissipation constant $\sigma$ result in different circularisation rates, and therefore the parameters of equation (\ref{eq:PcTAff}; or, more properly, its time derivative) ought to depend on $\sigma$. Using our fiducial calibration of $\sbp$, $\gamma = 0.22 \pm 0.01$ and $k_{P} = 1.9 \, \dif$ give the best least-squares fit. For H10's calibration, $\gamma$ remains effectively unchanged and $k_{P} = 1.5 \, \dif$; while for a planetary tide of $\sbp = 3 \times 10^{-6}$, $k_{P}$ increases to $2.6 \, \dif$ while $\gamma$ again remains nearly unchanged. These three cases are shown in Figure \ref{fig:PcAge}.

Empirically, we find that the value of $k_{P}$ scales with $\sbp$ in the same way as $\Pc$ with age; that is,
$$ k_{P} \propto \sbp^{\gamma}, $$
so that we may define an overall ``circularisation--age relation'' as
\begin{equation} \label{eq:PcAge}
    \Pc(t) \approx 1.9 \, \dif \left( \frac{\sbp}{8 \times 10^{-7}} \frac{t}{1 \, \Gyr} \right)^{2/9}.
\end{equation}
This relationship demonstrates explicitly the degeneracy between ages and dissipation strengths for HJs, which open clusters and other well-dated populations may be able to break.

The upshot of this section is this: that well-sampled populations of HJs at different ages can reveal the strength of tidal dissipation by way of the evolution of their circularisation periods, much as has been done for close binary stars by MM05 and H10.

\subsubsection{How precise are these constraints?} \label{s:PEA:precision}

Before we continue, let us consider the characteristic uncertainties of inferences about tidal dissipation made via this technique. Let the uncertainty of an estimate of $\Pc$ for some coeval population be denoted by $\delta \Pc$, and similarly $\delta \sbp$ and $\delta t$ for $\sbp$ and the population's age. By total differentiation of equation (\ref{eq:PcAge}), we find
$$ \delta \Pc \propto \gamma \sbp^{\gamma - 1} t^{\gamma} \, \delta \sbp + \gamma \sbp^{\gamma} t^{\gamma - 1} \, \delta t, $$
where we consider a general power-law index $\gamma$, rather than the value $\gamma \approx 2/9$ fit from our models. Division through by $\Pc$ gives
\begin{equation} \label{eq:dPc/Pc}
    \frac{\delta \Pc}{\Pc} = \gamma \left( \frac{\delta \sbp}{\sbp} + \frac{\delta t}{t} \right).
\end{equation}

There are two limiting cases of interest: First, that in which the population's age is known with great precision, so that $\delta t \to 0$. One derives
\begin{eqnarray} \label{eq:dsig/sig_Pc}
    \frac{\delta \sbp}{\sbp} &\approx& \frac{1}{\gamma} \frac{\delta \Pc}{\Pc}.
\end{eqnarray}
A possible application of this formula is when a population has few members, so that $\Pc$ is not well constrained. Alternatively, it may be used to estimate the range in which the circularisation period may fall when a previous estimate of $\sbp$ is assumed.

The other important application is when a population is well resolved, so that $\delta \Pc \to 0$. One derives
\begin{equation} \label{eq:dsig/sig_t}
    \frac{\delta \sbp}{\sbp} \approx \frac{\delta t}{t};
\end{equation}
this is merely a restatement of that which we have already argued: that the uncertainty of the inferred dissipation is determined primarily by that of the age.

\subsection{The Effect of Stellar Evolution} \label{s:PEA:SSE}

Tidal dissipation is highly sensitive to the radii of the dissipating bodies: referring to equation (\ref{eq:tauDef}), we see that a factor of $\sim 2$ in radius amounts to a factor of $\sim 10^{3}$ in the friction time-scale! Therefore, for planetary systems, stellar and planetary evolution may contribute significantly to orbital evolution. It is consequently of interest to consider whether our predictions for HJs around young stars are affected by these processes. We will focus primarily on stellar evolution (which is essentially unaffected by tidal dissipation), but the evolution of the planetary radius may also evolve significantly (e.g., \citealt{leconte+10}, H10, H12). We have repeated each simulation in this section with planetary radii of $1 \RJ$ and $1.4 \RJ$, in addition to the default $1.2 \RJ$. The effect of varying the planetary radius appears to be that the number of circularised planets changes relative to that of eccentric planets, with greater radii resulting in a greater proportion of circularised planets; however, other properties of the overall population, such as the circularisation period as a function of age, appear to be unaffected.

For solar-mass stars, significant radial evolution occurs during the pre- and post-main-sequence stages, which correspond to ages $t \la 30$--$50 \, \Myr$ and $t \ga 10 \, \Gyr$, respectively. Because we are most interested in youthful systems, we will further narrow our inquiry to pre-main-sequence contraction. During this period, predicting the fate of any HJs that may exist is nontrivial, as several effects determine the course of tidal evolution.

\subsubsection{Qualitative Considerations}

Let us consider a representative system consisting of a $1 \MJ$ planet orbiting a distance $0.03 \, \AU$ from a T Tauri star of $1 \MSol$ and $2 \RSol$. A reasonable criterion for significant orbital evolution during the pre--main sequence can be obtained by insisting that the tidal-friction time-scale due to stellar tides be comparable to the Kelvin--Helmholtz contraction time-scale of the star:
\begin{eqnarray}
    t_{\rm KH} &=& \frac{G M_{\star}^{2}}{R_{\star} L_{\star}}, \nonumber \\ &\approx& 12 \, \Myr \left( \frac{M_{\star}}{\MSol} \right)^{2} \left( \frac{R_{\star}}{2 \RSol} \right)^{-1} \left( \frac{L_{\star}}{L_{\sun}} \right)^{-1}. \label{eq:tKH}
\end{eqnarray}
By equating expressions (\ref{eq:tauDef}) and (\ref{eq:tKH}), we find that significant orbital evolution can be expected during the pre--main sequence if the semimajor axis of the orbit is less than
\begin{eqnarray}
    a_{\rm pre} &\approx& 0.02 \, \AU \left( \frac{\sbs}{10^{-7}} \right)^{1/8} \left( \frac{M_{\star}}{\MSol} \right)^{1/4} \nonumber \\ && \times \left( \frac{M_{\rm pl}}{\MJ} \right)^{1/8} \left( \frac{R_{\star}}{2 \RSol} \right)^{9/8} \left( \frac{L_{\star}}{L_{\sun}} \right)^{-1/8}.
\end{eqnarray}
At first blush, it would appear that our representative system with $a = 0.03 \, \AU$ is unlikely to evolve during the pre--main sequence. However, it is important to note that the efficiency of tidal dissipation in a pre-main-sequence star may be substantially greater than in a main-sequence star, owing to their fully convective structure and augmented radii. A more reasonable estimate for the stellar dissipation constant might then be the value $\sbs \sim 10^{-5}$ calculated for fully convective, main-sequence M-type stars (H12). In that case, the time-scale of circular-orbit evolution at $0.03 \, \AU$ would be a mere $1.6 \, \Myr$, comparable to the ages of the candidate systems and to the timescale of the host star's contraction; for such systems, we might indeed expect the planet to migrate tidally during the pre--main sequence. Doubling the size of the orbit increases this to $\sim 400 \, \Myr$, which is much longer than the stellar contraction time-scale.

Additionally, we must remember that the direction of tidal migration (inward or outward) is determined by the relative angular velocities of the orbit and the stellar spin. Planets with orbital periods shorter than the stellar spin period will spiral into their hosts, while those with longer orbital periods will migrate outward. The critical semimajor axis at which the orbit and spin are synchronous (the so-called corotation radius) is given by Kepler's third law:
\begin{equation} \label{eq:aSync}
    a_{\rm co} \approx 0.04 \, \AU \left( \frac{M_{\star}}{\MSol} \right)^{1/3} \left( \frac{P_{\rm \star, TTS}}{3 \, \dif} \right)^{2/3},
\end{equation}
where $P_{\rm \star, TTS}$ is the stellar spin period as a T Tauri star.

As the host contracts onto the main sequence, its spin period shortens due to conservation of angular momentum, decreasing both $a_{\rm co}$ and $a_{\rm pre}$ accordingly. This suggests that the potential for planets to move in- or outwards under tidal torques diminishes quickly with time; we therefore would expect the distribution of orbital periods among young HJ systems to be affected significantly only if (1) a significant fraction of these planets arrive early in the star's pre-main-sequence evolution and (2) these planets arrive on orbits with periods shorter than the stellar spin period.

If this be the case, then we would expect the following main effects on the period distribution of young HJs: Firstly, orbital periods up to that corresponding to an average value of $a_{\rm co}$ for the primordial HJ population would be largely devoid of planets, since these would have been destroyed through in-spiral. Secondly, those which remain would accumulate roughly between this same value of $a_{\rm co}$ and a corresponding value of $a_{\rm pre}$. Finally, over time, stellar tides during the main sequence would gradually drag planets inward as the stellar spin period lengthened under magnetic braking, eventually re-populating semimajor axes down to $\sim a_{\rm Ro}$.

\subsubsection{Numerical Exploration}

\begin{figure}
    \includegraphics[width=\columnwidth]{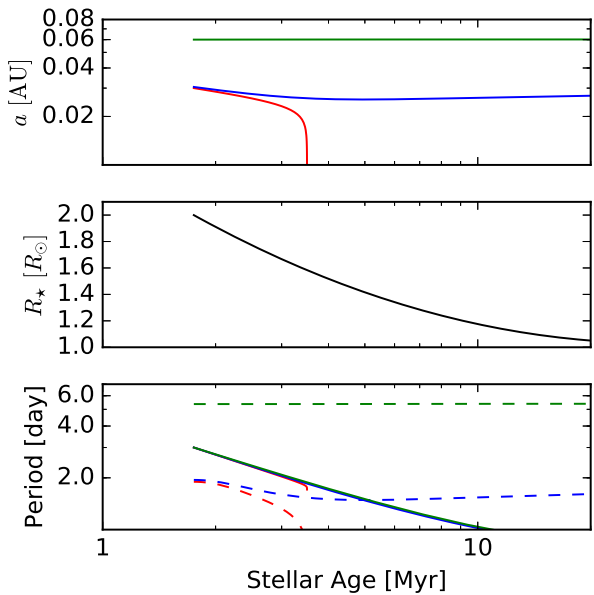}
    \caption{Tidal evolution of a $1 \MJ$ planet around a pre-main-sequence host with $M_{\star} = 1 \MSol$ and $\sbs = 10^{-5}$. Depicted are the semimajor axes (top panel) and stellar spin periods (bottom, solid contours) of systems characteristic of the three predominant outcomes of our scenario, for initial semimajor axes of $0.03 \, \AU$ (red), $0.0305 \, \AU$ (blue), and $0.06 \, \AU$ (green) with zero eccentricity. The evolution of the stellar radius (middle) follows the models of \citet{bhac15}. The evolution of the stellar spin is dominated by the contraction of the star rather than by the infusion of angular momentum from the planet's orbit. Notice that in marginal cases (blue), the direction of the planet's migration can be reversed as the stellar spin outstrips the planet's orbit (bottom, dashed contours).}
    \label{fig:PMSHJ}
\end{figure}

To evaluate the validity of these predictions, we can simply repeat some of our previous numerical experiments, this time including models of pre-main-sequence and main-sequence stellar evolution by \citet{bhac15}. As a first pass at this, we examine the archetypal system considered in the previous section. The evolution of three realisations of this system is illustrated in Figure \ref{fig:PMSHJ}. The most notable feature of these case studies is their illustration of the extreme sensitivity of the outcome to the initial semimajor axis of the planet: a difference of just a few parts in $10^{4}$ in the initial semimajor axis determines that the planet inserted at $0.03 \, \AU$ will not survive the pre--main sequence of its star, whereas that at $0.0305 \, \AU$ will. Although the transition between in- and out-spiral does not occur at $a_{\rm co}$ as predicted by equation (\ref{eq:aSync}), the qualitative picture described in the preceding subsection appears to be correct.

We now examine our qualitative predictions about the effect of pre-main-sequence evolution on populations of coeval planets. We initialise three ensembles, one each in which all planets are inserted into their orbits at protostellar ages of $1 \, \Myr$, $3 \, \Myr$, and $10 \, \Myr$; we then observe the state of each ensembles after aging by a factor of $\sim 3$ successively until an age of $11 \, \Gyr$ (note that solar-mass stars arrive on the main sequence at an age of $\sim 50 \, \Myr$). The initial protostellar spin periods are chosen such that, absent the effects of tides, the protostar would spin once every $3$ days when its radius reached $2 \RSol$; after the protostar reaches the main sequence (which we define by the radius contracting to $1 \RSol$), we include the effects of magnetic braking on the stellar spin following \citet{bo09}. We also note that, because $t_{\rm F,\star}$ is far more sensitive to $R_{\star}$ than to $\sigma_{\star}$, our choice of $\sbs$ for this experiment makes very little difference to the outcome of pre-main-sequence evolution (although it certainly matters for the long-term evolution of these ensembles during the main sequence); we have chosen $\sbs = 10^{-7}$ for consistency with our previous experiments.

\begin{figure*}
    \includegraphics[width=\columnwidth]{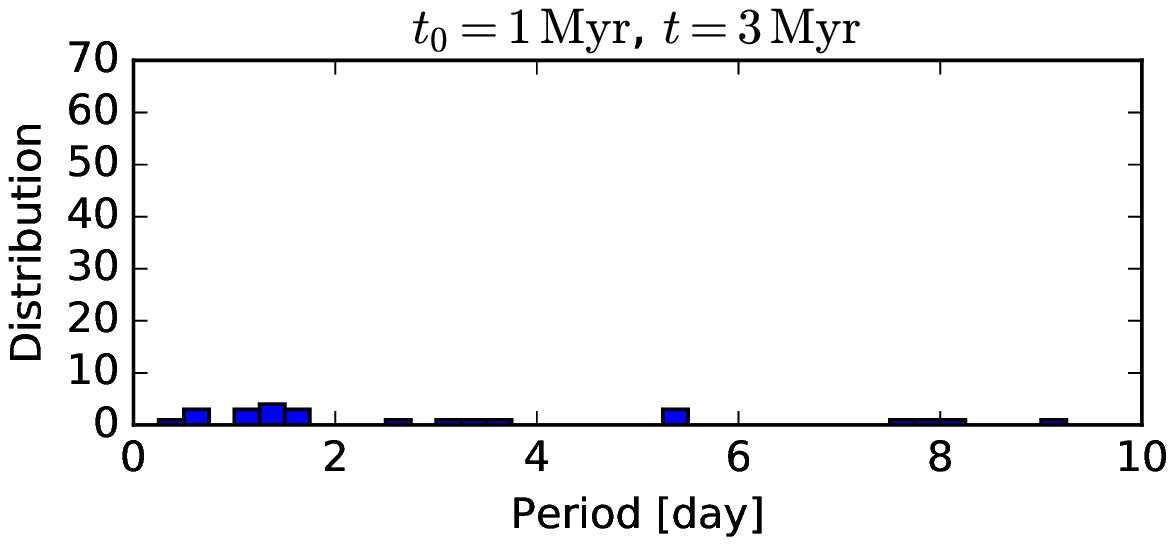}
    \includegraphics[width=\columnwidth]{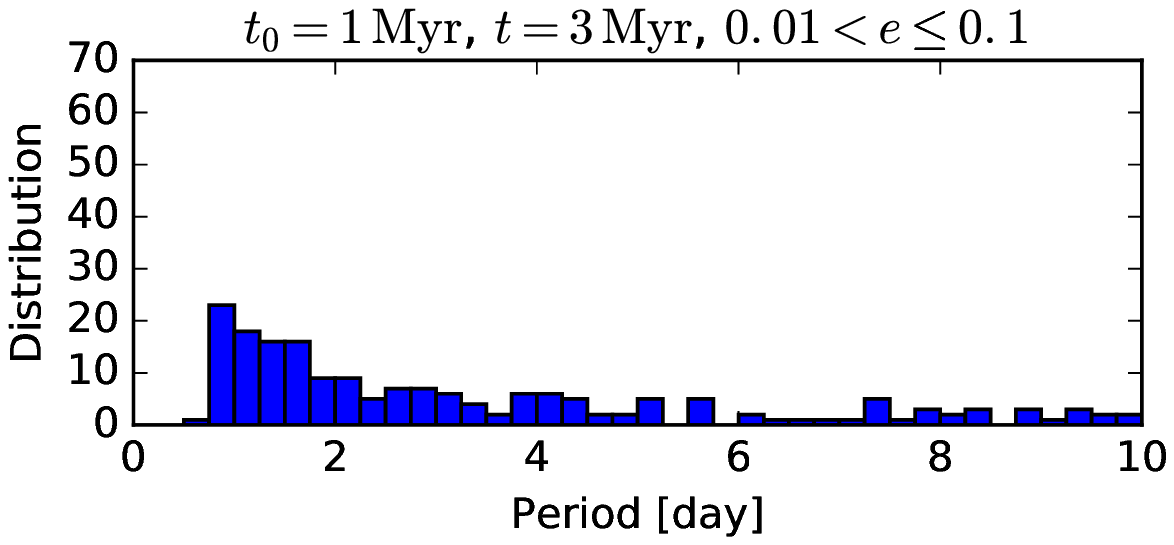}
    \includegraphics[width=\columnwidth]{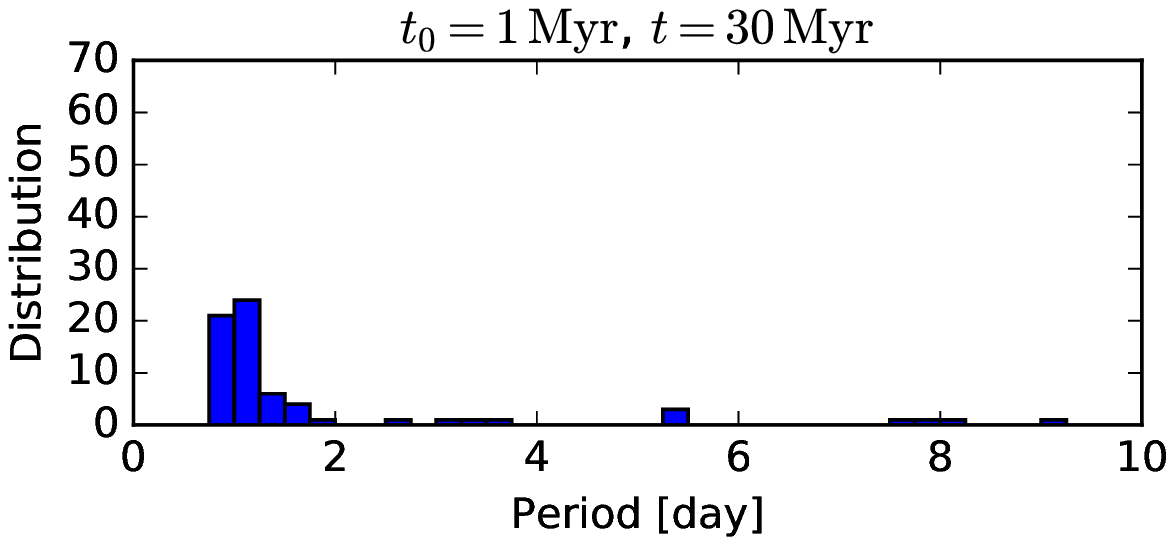}
    \includegraphics[width=\columnwidth]{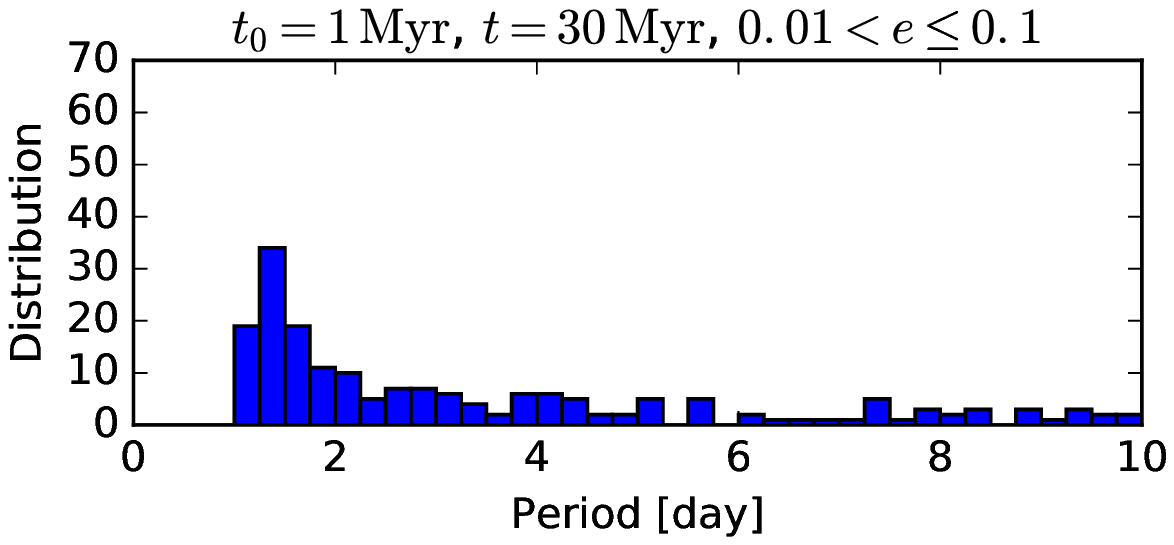}
    \includegraphics[width=\columnwidth]{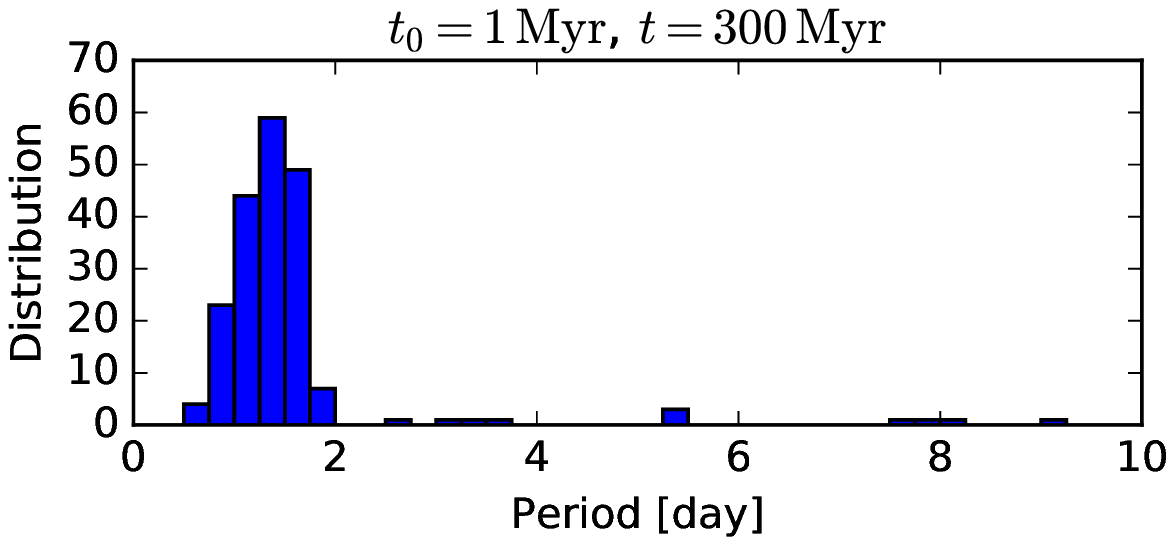}
    \includegraphics[width=\columnwidth]{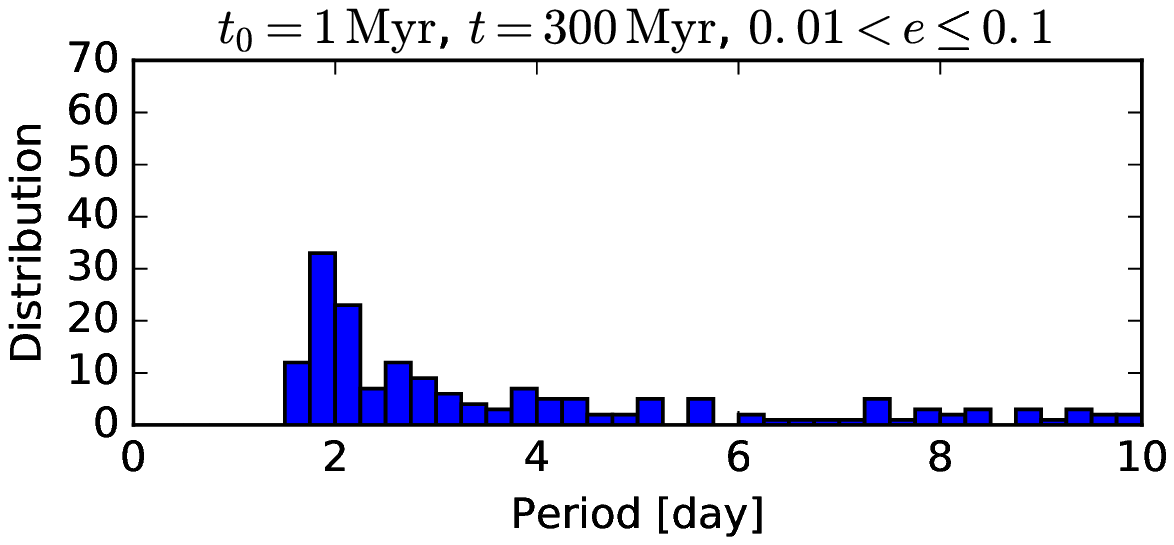}
    \includegraphics[width=\columnwidth]{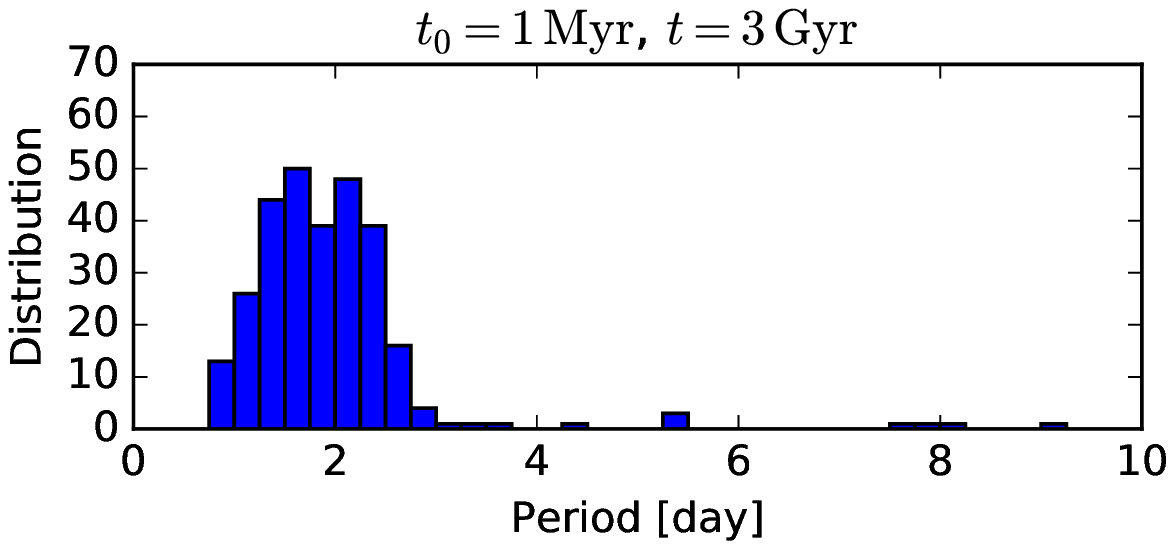}
    \includegraphics[width=\columnwidth]{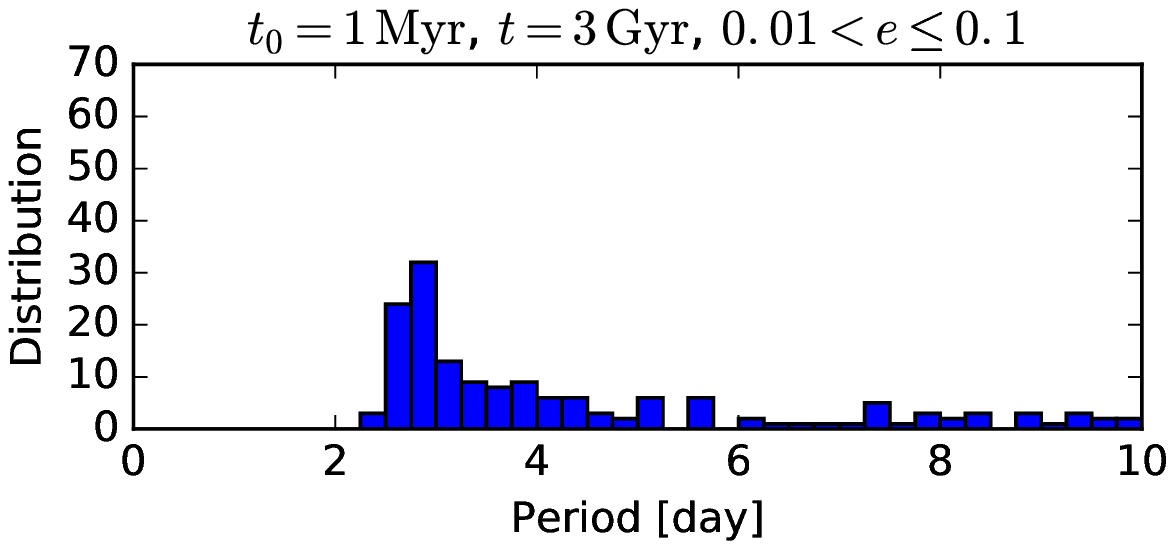}
    \includegraphics[width=\columnwidth]{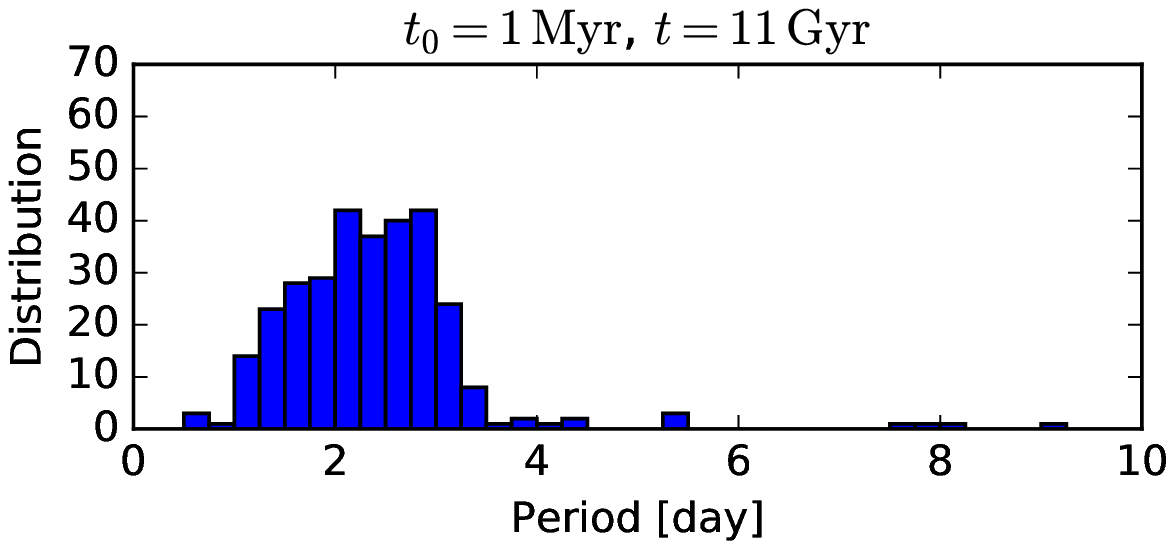}
    \includegraphics[width=\columnwidth]{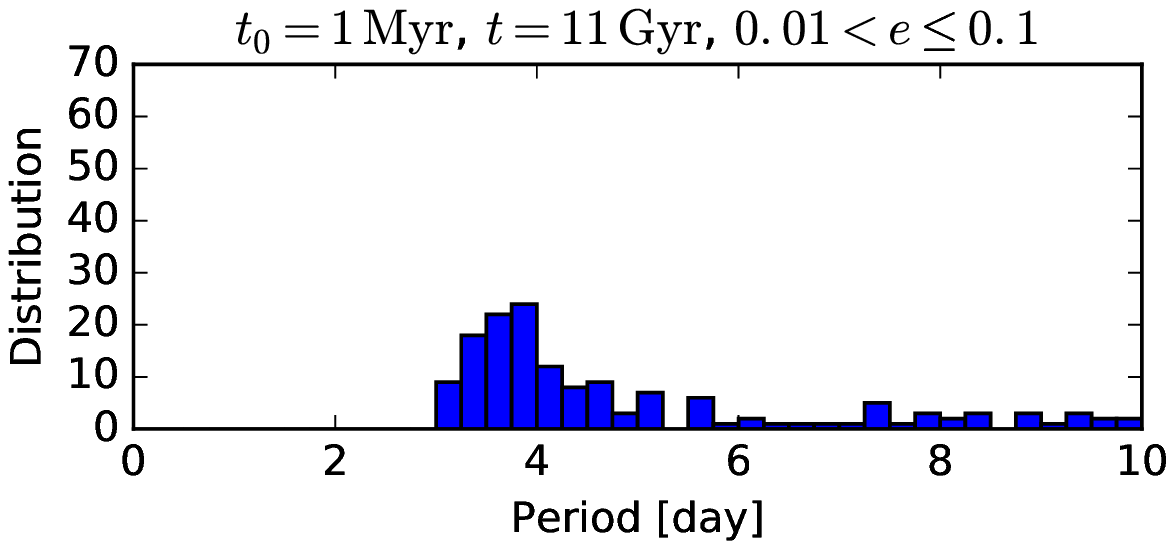}
    \caption{Histograms of the orbital periods of our model planets, now taking into account the evolution of the stellar radius and spin. Systems were initialised at a stellar age of $1 \, \Myr$ using the same period and eccentricity distribution as in Figure \ref{fig:PHist}. The left-hand column displays planets with $e \leq 0.01$ and the right with $0.01 < e \leq 0.1$.}
    \label{fig:PHist+PMSHJ}
\end{figure*}

In Figure \ref{fig:PHist+PMSHJ}, we show the evolution of the period distribution of the ensemble initialised at an age of $1 \, \Myr$. In comparison to Figure \ref{fig:PHist}, we see that the models accounting for stellar evolution produce fewer surviving HJs at young ages, particularly for systems still on the pre--main sequence ($t \la 50 \, \Myr$). We attribute this to the in-spiral of planets deposited on extremely short-period orbits early in protostellar contraction. After the host star reaches the main sequence ($t \ga 50 \, \Myr$), the evolution of the model planets is much the same as before: the piles-up have essentially the same shapes and scales at each age and in both eccentricity ranges. On the whole, then, the basic properties of the HJ population at each age is insensitive to stellar evolution on the pre--main sequence; the largest change is the relative number of HJs present in the population at periods less than $\sim 1$ day for systems less than $\sim 50 \, \Myr$ old. We do not find that the simulations initialised at $3 \, \Myr$ and $10 \, \Myr$ yield significantly different results than that at $1 \, \Myr$.

\subsubsection{Model Uncertainties}

We concede that our investigation of pre-main-sequence evolution is incomplete in that we have taken only equilibrium tides into account. Recent theoretical studies have shown that dissipation by dynamical tides may be several orders of magnitude more efficient than that by equilibrium tides for pre-main-sequence host stars \citep{bm16,gallet+17}. \citeauthor{bm16} find that dynamical tides can significantly alter the qualitative behavior of young HJ systems: in particular, they find that the migration of planets on very short orbital periods can reverse its direction relative to migration under the equilibrium tide alone, so that planets move outward instead of spiralling into their stars. If these models are a more realistic representation of pre-main-sequence evolution, then one would expect young HJs to orbits systematically farther from their stars than older planets. Once again, however, we emphasize that this would require these planets to arrive within the first $\sim 10 \, \Myr$ of stellar evolution.

\section{Discussion} \label{s:disc}

\subsection{Two Practical Examples} \label{s:disc:ex}

\begin{table*}
    \centering
    \begin{tabular}{lllllll}
        \hline
        Cluster & $t~[\Gyr]$ & Planet & $P~[\dif]$ & $e$ & $M_{\rm pl} \sin \iota~[\MJ]$ & Reference \\
        \hline
        Hyades/Praesepe & $0.7 \pm 0.1$ & Pr0201\,b & $4.426$ & $ 0 \, (< 0.2)$ & $0.54$ & 1 \\
        & & Pr0211\,b & $2.146$ & $0.017 \pm 0.010$ & $1.88$ & 1, 5 \\
        & & HD\,285507\,b & $6.088$ & $0.086 \pm 0.018$ & $0.92$ & 2 \\
        \hline
        M\,67 & $4.0 \pm 0.5$ & YBP401\,b & $4.087$ & $0.16 \pm 0.08$ & $0.42$ & 4, 6 \\
        & & YBP1194\,b & $6.960$ & $0.31 \pm 0.08$ & $0.33$ & 3, 6 \\
        & & YBP1514\,b & $5.118$ & $0.27 \pm 0.09$ & $0.40$ & 3, 6 \\
        \hline
    \end{tabular}
    \caption{Parameters of HJs in Open Clusters. References -- 1: \citet{quinn+12}, 2: \citet{quinn+14}, 3: \citet{brucalassi+14}, 4: \citet{brucalassi+16}, 5: \citet{malavolta+16}, 6: \citet{brucalassi+17}}
    \label{tab:OCPlanetData}
\end{table*}

\begin{table*}
    \centering
    \begin{tabular}{lllllll}
        \hline
        Cluster & $t~[\Gyr]$ & Star & $M_{\star}~[\MSol]$ & $R_{\star}~[\RSol]$ & $P_{\rm rot}~[\dif]$ & Reference \\
        \hline
        Hyades/Praesepe & $0.7 \pm 0.1$ & Pr0201 & $1.234$ & $1.167$ & $5.63$ & 2, 5 \\
        & & Pr0211 & $0.952$ & $0.868$ & $7.97$ & 2, 5, 6 \\
        & & HD\,285507 & $0.734$ & $0.656$ & $11.98$ & 1, 4 \\
        \hline
        M\,67 & $4.0 \pm 0.5$ & YBP401 & $1.14$ & $1.25$ & -- & 7, 8 \\
        & & YBP1194 & $1.01$ & $0.99$ & -- & 3, 7, 8 \\
        & & YBP1514 & $0.96$ & $0.89$ & -- & 3, 7, 8 \\
        \hline
    \end{tabular}
    \caption{Parameters of HJ Host Stars in Open Clusters. References -- 1: \citet{delorme+11}, 2: \citet{quinn+12}, 3: \citet{brucalassi+14}, 4: \citet{quinn+14}, 5: \citet{kovacs+14}, 6: \citet{malavolta+16}, 7: \citet{brucalassi+16}, 8: \citet{brucalassi+17}}
    \label{tab:OCStarData}
\end{table*}

\begin{figure*}
    \includegraphics[width=\columnwidth]{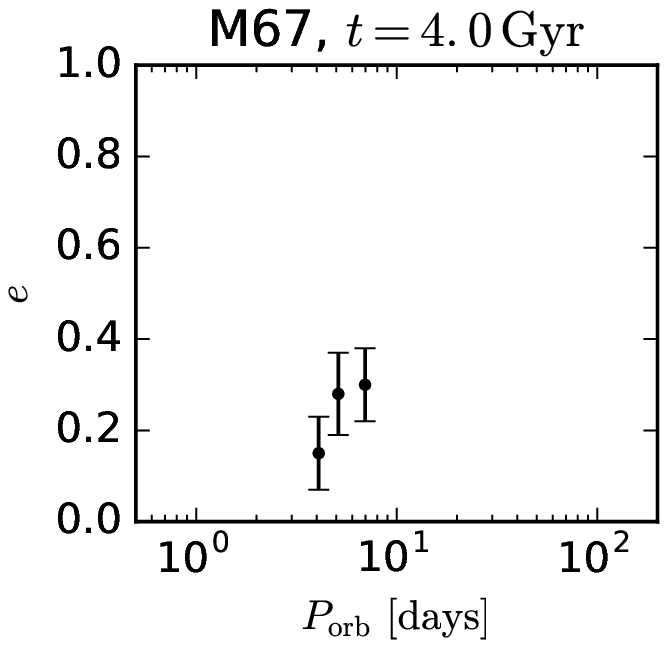}
    \includegraphics[width=\columnwidth]{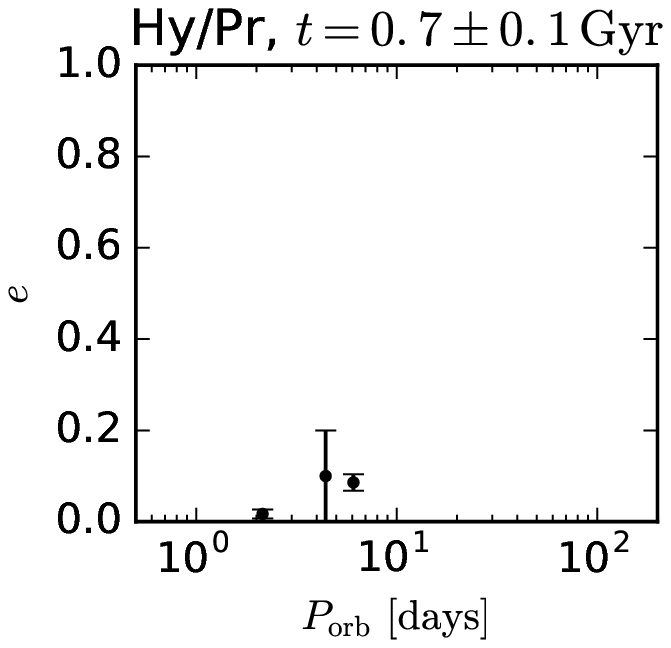}
    \caption{$P$--$e$ diagrams of the HJ populations in the clusters M\,67 (upper panel) and Hyades/Praesepe (lower panel).}
\end{figure*}

In the following two subsections, we demonstrate both the utility and the limitations of the current sample of cluster-member HJs in constraining tidal dissipation. The two populations we consider are too small as yet for the methods discussed in Section \ref{s:PEA} to be applicable; but, none the less, they reveal the latent potential in such populations for more meaningful constraints in the future. The parameters of the open-cluster systems we consider are listed in Tables \ref{tab:OCPlanetData} and \ref{tab:OCStarData}.

\subsubsection{M67} \label{s:disc:ex:M67}

Recently, M\,67 was the subject of a seven-year campaign of radial-velocity monitoring in search of substellar objects \citep{pasquini+12,brucalassi+14,brucalassi+16,brucalassi+17}. This has revealed five jovian planets in the cluster, including three HJs. Most fortuitously for the purposes of this study, all three display moderately eccentric orbits ($e \ga 0.1$). To boot, the solar-like age of M\,67, some $4.0 \pm 0.5 \, \Gyr$ \citep[etc.]{barnes+16,stello+16}, allows a direct comparison of these planets to HJs in the field.

Due to the small sample size in this cluster, we have applied the calibration technique used by H10 to constrain $\sbp$ for the planets in this cluster. For each system, we generate a large model population of planets identical to that observed around a corresponding star. We evolve these planets to the minimum and maximum estimated ages of the systems (in this case, the age of M\,67). To constrain the value of $\sbp$, we adjust this parameter in such a way that at the minimum (maximum) age of the system, the model planets are consistent with the smallest (largest) eccentricity within the error bar of the real planet's measured eccentricity. This allows us to place an upper (lower) bound on $\sbp$ for each planet.

By this procedure, we find the following constraints on the dissipation constants for each of the three known HJs in M\,67. For YBP401\,b \citep{brucalassi+16},
$$ 9.0 \times 10^{-7} < \sbp < 3.5 \times 10^{-6}; $$
for YBP1194\,b \citep{brucalassi+14},
$$ 3.0 \times 10^{-6} < \sbp < 1.5 \times 10^{-5}; $$
and for YBP1514\,b \citep{brucalassi+14},
$$ 1.5 \times 10^{-6} < \sbp < 5.5 \times 10^{-6}. $$
Altogether, then, we find that the orbits of the HJs in M\,67 are consistent with a single planetary dissipation constant $\sbp = (3 \pm 1) \times 10^{-6}$. On the basis of this constraint and our evolutionary models, we predict that, if additional HJs are discovered in this cluster, their position in the $P$--$e$ diagram will be consistent with a circularisation period of $\approx 3.7 \, \dif$, with a margin of error of $\approx 0.4 \, \dif$ (equation \ref{eq:dPc/Pc}).

The inferred value is about an order of magnitude greater than the dissipation inferred by H10 using the same technique, but on par with H12's calculations from models of planetary interiors. Again, an order-of-magnitude offset is also understandable in terms of the degeneracy between planetary mass and inclination for systems such as these known only from radial-velocity observations.

It is somewhat surprising that the M\,67 HJs, few though they may be, can none the less yield such a tight constraint on the dissipation in these planets. Even if our simplified model leads us to under-estimate the uncertainty of the inferred values, an order-of-magnitude error bar still would give a result comparable in precision to constraints based on larger sample sizes (H10, B17) or more sophisticated techniques \citep{penev+16}. In part, this could be due to the lucky chance that M\,67's planets happen to have eccentric orbits at the time of their discovery; this, coupled with the possibility that M\,67's stars host HJs more frequently than stars in the field \citep{brucalassi+16,brucalassi+17}, might suggest that high-eccentricity migration is the dominant channel for producing these planets in old star clusters \citep*[perhaps due to more frequent stellar encounters;][]{shara+16}. On the whole, we interpret this example as a ``best-case scenario'' for the extraction of constraints on tidal dissipation.

\subsubsection{Hyades and Praesepe} \label{s:disc:ex:HyPr}

The coeval clusters Praesepe and the Hyades host at least three HJs between them: Pr0201\,b and Pr0211\,b in Praesepe \citep{quinn+12} and HD\,285507\,b \citep{quinn+14} in the Hyades. Until recently, the age of these clusters was considered to be well constrained at $\sim 600 \, \Myr$ \citep[][etc.]{delorme+11}. However, \citet{bh15a,bh15b} report that both clusters can be described well by an age of $\sim 800 \, \Myr$ when one accounts for the effects of stellar rotation. We therefore allow for an uncertainty of $\sim 200 \, \Myr$ in the age of the Hyades/Praesepe HJs between these competing estimates.

The eccentricity of Pr0201\,b is not well constrained. Radial velocity measurements indicate $e < 0.2$, but the planet's low projected mass makes a more precise measurement difficult; as such, the authors of the discovery paper favor a circular orbit. Even the more lenient constraint requires $\sbp \sim 10^{-5}$ in order to sufficiently circularise all orbits at this period and system age, an order of magnitude stronger than the tidal strength inferred either in the field or in M\,67. HD\,285507\,b, which has a larger orbit with $e = 0.086 \pm 0.018$, gives an even more extreme result.

The shortest-period planet between the two clusters in Pr0211\,b \citep{quinn+12}, at $P = 2.146$ days. Its eccentricity is consistent with zero, $e = 0.017 \pm 0.010$, per \citet{malavolta+16}. \citeauthor{malavolta+16} also discovered Pr0211\,c, a long-period ($P \ga 3500$ days), eccentric ($ e \ga 0.6$), massive ($M_{\rm p} \sin \iota \approx 8 \MJ$) outer planet. The architecture of this system is reminiscent of a high-eccentricity history by planet--planet scattering \citep[e.g.,][]{rf96,wm96} or perhaps the Kozai--Lidov mechanism \citep[e.g.,][]{wm03,ft07,naoz+11}. Given that we have so far considered only systems with a single known planet, it is reasonable to ask whether our simplistic model be adequate to describe the circularisation of Pr0211\,b, in light of additional eccentricity pumping that may have occurred during migration. However, once the planet approaches the star at distances sufficiently small to trigger tidal migration, relativistic precession becomes sufficient to decouple the inner planet from the secular influence of the outer \citep[e.g.,][]{ft07,naoz+13}; thus, the tidal evolution of Pr0211\,b should still be described reasonably well by our model. We find that its present orbit is roughly consistent with our result for the transiting sample, $\sbp = 5 \times 10^{-6}$ being sufficient to circularise the planet from a Roche-grazing orbit within $600 \, \Myr$ and $\sbp = 3 \times 10^{-6}$ within $800 \, \Myr$. A more detailed study of this system's dynamical history may be in order, if a proper constraint on tidal dissipation is to be obtained.

The HJs of Praesepe and the Hyades do not yield a picture of tidal circularization that is consistent with their older counterparts in M\,67 and in the field. In particular, the circularity of Pr0210\,b and HD\,285507\,b requires a higher value of $\sbp$ in order to be consistent with high-eccentricity migration. Of course, given the small number of known systems in these clusters, it is entirely possible that some of them formed with circular orbits and some eccentric. All in all, these two clusters provide a natural example of the difficulties that remain for the study of HJ formation, even with the benefit of a well-constrained age.

\subsection{Pre-Main-Sequence Systems} \label{s:disc:young}

In Section \ref{s:PEA:SSE}, we discussed the possible effects of pre-main-sequence stellar evolution on populations of HJs or proto-HJs. As of this report, four HJ candidates have been reported around T Tauri stars by \citet{vaneyken+12}, \citet{donati+16}, \citet{johnskrull+16}, and \citet{yu+17}. The planetary statuses of these candidates remain somewhat unclear. None the less, it is interesting to consider the implications of these systems, should they prove to be genuine.

The first, most obvious conclusion is that disc-driven migration is likely a viable pipeline for producing close-in massive planets. Additional inferences about the exact nature of planetary migration and tidal interactions must wait until additional systems can be discovered at these ages. Still, it is curious that three of these systems orbit well beyond the pile-up observed among main-sequence systems; should this trend continue, it would be consistent with predictions following \citet{bm16} that young HJs tend to migrate outward from extremely short-period orbits during pre-main-sequence evolution. If so, the age at which HJs begin to occur on orbits closer than $\sim 0.05 \, \AU$ might provide important information about the contribution of high-eccentricity migration.

\subsection{Future Prospects} \label{s:disc:future}

Clearly, one of the largest limits in the study of HJ formation and dynamics is the number of systems known and the extent of our ability to characterize systems in detail. This will be as true for planets in star clusters as it is for those in the field. However, there is reason to hope that our knowledge may be on track to improve drastically in the near future.

The {\it Gaia} \citep{gaia+16} and {\it TESS} \citep{ricker+15} missions may aid in the discovery of planetary systems in open clusters. Although {\it Gaia} was not built specifically for the detection of transiting planets, \citet{dz12} estimate an incidental yield of a few hundred HJs. {\it Gaia}'s contribution to our knowledge of exoplanets will likely be greatest for long-period planets \citep{perryman+14}; this presents the possibility that the satellite will reveal exterior companions to HJs, such as Pr0211\,c. A more exciting prospect is that {\it Gaia} may be able to discover highly eccentric planets on their way to becoming HJs via high-eccentricity migration, which have thus far been elusive \citep*{dawson+15}. Such discoveries reveal valuable information on the formation of these planets, even if they do not reveal tidal dissipation directly.

{\it TESS}, meanwhile, is expected to reveal legions of HJs across the sky. One would hope that, between the two, some of the {\it Gaia} or {\it TESS} planets might be found in star clusters. If so, they will be natural targets for spectroscopic and/or asteroseismic followup, orbit characterization, and long-term monitoring. Hopefully, this would lead ultimately to a $P$--$e$ diagram comparable in completeness to that available for planets in the field.

On the other hand, FGK stars in the majority of galactic clusters may be too faint for either detection in transit surveys or long-term followup. In that case, the more promising pipeline for HJ discoveries may be long-term spectroscopic and photometric monitoring of stars in nearby clusters. Several such campaigns are either underway or planned (e.g., \citealt{malavolta+16}, \citealt{quinn16}).

\subsection{Extended Applications}

We have so far confined our discussion to planetary systems in clusters, since these are explicitly coeval and their ages are reliably determined. However, the foregoing might apply equally well to systems in the field, provided that their ages be reliably estimated. Until recently, this has not been considered possible, owing to the imprecision of common techniques such as isochrone-fitting. However, ongoing advancements in gyrochronology and asteroseismology present the possibility of better and more widespread age estimates for these stars. Soon, it may be possible to resolve large, roughly coeval populations of HJs around stars in the field. For this to be successful, various theoretical uncertainties must be addressed regarding the applicability and effectiveness of these techniques (particularly gyrochronology) to both single stars \citep{vansaders+16} and HJ hosts (\citealt*{maxted+15}; \citealt{penev+16}; and references therein).

\section{Conclusions} \label{s:conc}

We have investigated the extent to which HJs discovered in star clusters, by virtue of their well-constrained ages, may be able to illuminate the formation of these planets, concentrating on the constraints that they may yield on tidal dissipation in these planets and their host stars. To do so, we have studied the properties and evolution of a homogeneous, coeval population of giant planets under equilibrium tides.

Our principal findings are the following:
\begin{enumerate}
    \item For an equilibrium tide with a dissipation efficiency consistent with planets in the field, there is a distinct possibility that highly eccentric HJs may be discovered around young stars ($\la 1 \, \Gyr$) with orbital periods less than $5 \, \dif$. Thus, the $P$--$e$ distribution of young planets may be able to disentangle populations of HJs that result from high- and low-eccentricity formation mechanisms.
    \item We find a relationship between the circularisation period of a coeval planet population, the age of that population, and the planetary dissipation constant $\sbp$, which takes the form of a power law (equation \ref{eq:PcAge}). This relationship applies best to populations for which the transition between circular and eccentric orbits can be identified. However, it can be applied cautiously to a population of planets lacking this transition, with correspondingly less certain results (equation \ref{eq:dPc/Pc} and its limiting cases). Our two most useful results on the uncertainties of these constraints are that (1) $\delta \sbp \propto \delta \Pc / \gamma$ in the limit where the age is well-known and (2) $\delta \sbp \propto \delta t$ in the limit where the population is well resolved.
    \item In the limit of the equilibrium tide, the evolution of HJ host stars on the pre--main sequence affects only weakly the properties of a coeval population of HJs; this is because the timescale of stellar contraction onto the main sequence is shorter than the tidal circularisation or migration time-scales for most planetary systems. The largest effect of pre-main-sequence stellar evolution is a depletion of HJ numbers on very short orbital periods relative to longer periods, but only for systems where the planet migrated within the first $\sim 10 \, \Myr$ of the protostellar lifetime. This effect may persist to ages of a few hundreds of $\Myr$, which are typical for galactic open clusters.
    \item The existing sample of HJs in clusters can yield constraints on tidal dissipation comparable in precision to those from the much larger population of planets in the field. This is largely because the known HJs in M\,67 have eccentric orbits. In the Hyades and Praesepe, the constraints are less clear, due to the small or unconstrained eccenticities of these planets. These two examples illustrate the importance and desirability of a robust sample of HJs in multiple clusters.
\end{enumerate}

\section*{Acknowledgements}

CEO is grateful for the generous support of the Litton Endowment via the Undergraduate Research Scholars Program at UCLA. He also thanks Jon Zink for thorough comments on an early incarnation of the manuscript and Alexander Stephan for computational advice. This work has made use of NASA's Astrophysics Data System and of the Exoplanet Orbit Database and the Exoplanet Data Explorer at {\tt exoplanets.org} \citep{han+14}.

\bibliographystyle{mnras}
\bibliography{tides_bib}

\appendix

\section{Calibration of the Tidal Theory} \label{s:field}

\subsection{Ensemble Calibration} \label{s:field:ensemble}

\begin{figure*}
    \includegraphics[width=2.0\columnwidth]{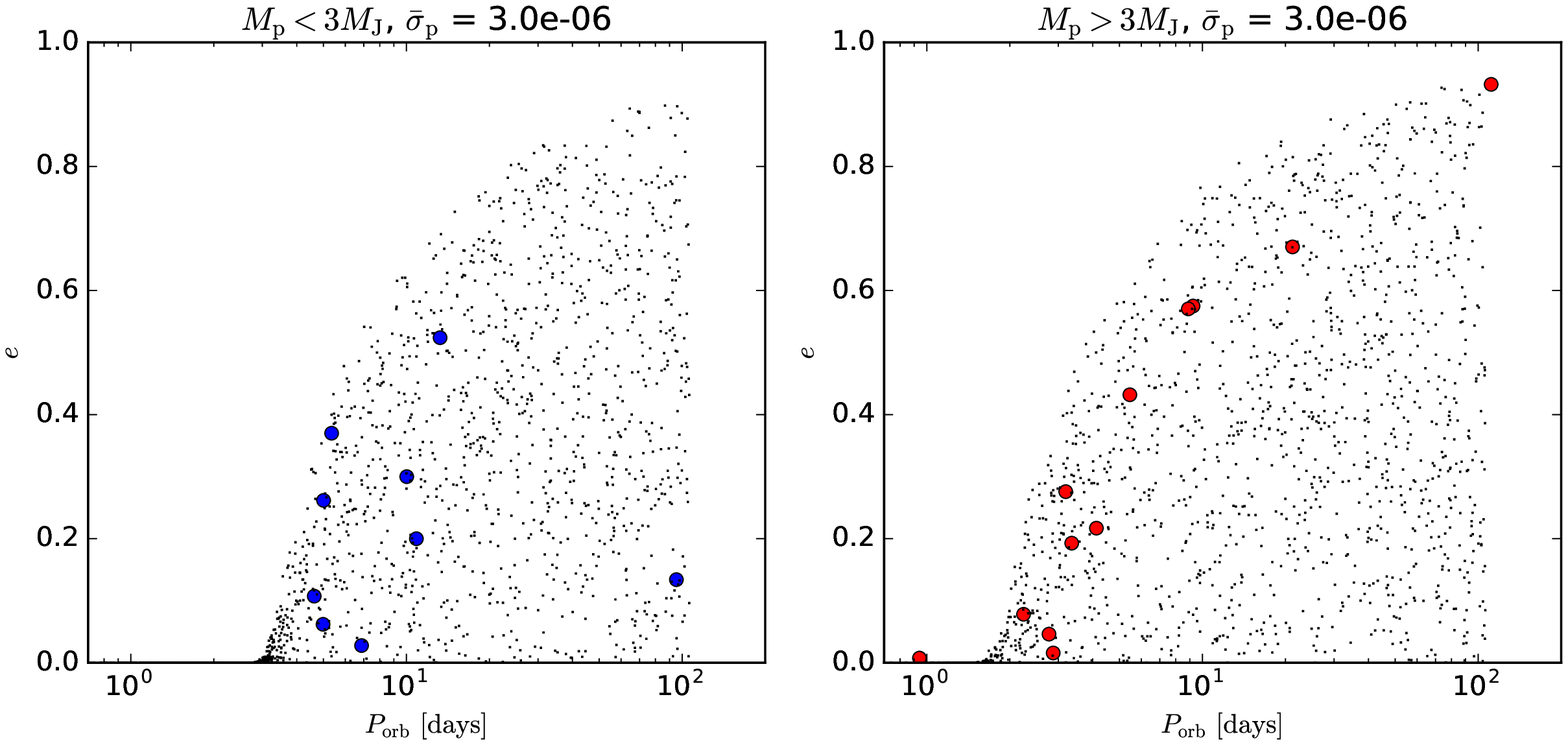}
    \caption{Period--eccentricity diagrams for low- (left panel) and high-mass (right panel) planets with nonzero eccentricities constrained by B17. Model planets are represented by single pixels and observed planets by colored dots. Both distributions are described fairly well by the values $\sbp = 3 \times 10^{-6}$ and $\sbs = 8 \times 10^{-8}$.}
    \label{fig:B17_fit}
\end{figure*}

As in H10, we begin our calibration of the planetary dissipation by examining the distribution of orbital solutions for an ensemble of jovian planets. We select the sample of transiting giant planets studied by B17, the advantages of which are twofold: Firstly, transiting giant planets by their very nature have unambiguous measurements of both masses and radii for planets and their host stars. Radii are especially important, as the tidal friction time-scale is highly sensitive to the radius ($t_{\rm F} \propto R^{-10}$). Additionally, B17 have improved upon existing eccentricity measurements for several systems, which limits the uncertainty in our calibration.

To carry out this preliminary calibration, we first divide the sample into two mass ranges: those with masses between $0.3 \MJ$ and $3 \MJ$ (hereafter ``low-mass planets''), and those more massive than $3 \MJ$ (``high-mass planets''). This division roughly separates systems in which stellar tides are negligible from those in which they are not (although we will include stellar tides in all calculations).

In each population, we computed the mean planetary mass and radius: for low-mass planets, these were roughly $1 \MJ$ and $1.2 \RJ$; for high-mass planets, $7 \MJ$ and $1.1 \RJ$. These parameters were adopted as representative values for each population. We generated two populations of identical planets, one each for the low- and high-mass values. The orbital elements of the model systems were drawn from uniform distributions in $\log a$ and $e$. We considered all stellar hosts to be $3.0 \, \Gyr$-old stars of $1 \MSol$ and $1 \RSol$; we set their initial spin periods to be $30$ days and did not consider any spin--orbit misalignment. Planets were assumed to rotate pseudo-synchronously according to their eccentricities \citep{hut81}.

Figure \ref{fig:B17_fit} shows that both the low- and high-mass samples can be described broadly by a planetary dissipation constant of $\sbp = 3.0 \times 10^{-6}$. We adopt this as our starting point for the next step, in which we will refine the calibration using observed systems near the upper envelope of the model population.

\subsection{Individual Systems} \label{s:field:indiv}

We now will attempt to refine our calibration of the planetary dissipation constant by comparing several individual systems to ensembles of specific models for those systems (that is, which use the observed planetary and stellar parameters of each system, rather than average values). To do so, we assume that each individual system lies on the $P$--$e$ envelope of the best-fitting model at each age. This is a strong assumption; as we shall see, it does not hold true unless we wish to abandon the weaker assumption that all jovian planets should have similar dissipation constants under the equilibrium tide.

\subsubsection{CoRoT-16} \label{s:field:indiv:CoRoT16}

At the upper-left-most extreme of Figure \ref{fig:B17_fit} lies CoRoT-16b \citep{ollivier+12}, a $0.53 \MJ / 1.17 \RJ$ planet on a $5.35$-day orbit with $e = 0.37 \pm 0.12$ (B17). The stellar age of $6.7 \pm 2.8 \, \Gyr$ is sufficiently old that its $1.1 \MSol / 1.2 \RSol$ host star likely has evolved significantly since tidal interactions began. However, the planet's (relatively) low mass and long orbit suggest that tides raised on the star likely have not yet become significant in its orbital evolution.

By using the $1\sigma$ lower limits on the eccentricity and the stellar age, we obtain an upper limit of $\sbp = 1.0 \times 10^{-6}$ on the dissipation constant; similarly, the upper limits imply a lower bound of $\sbp = 3.4 \times 10^{-7}$.

\subsubsection{HAT-P-14} \label{s:field:indiv:HATP14}

At first glance, the orbit of HAT-P-14b \citep{torres+10} seems to lie close to the envelope implied by CoRoT-16b in the ensemble $P$--$e$ diagram. However, this planet's relatively high mass ($2.3 \MJ$) hinders significantly dissipation in the planet. In order to circularise all model planets with its $4.6$-day orbit to the measured eccentricity ($0.107$, with negligible uncertainty for our purposes) within the minimum age of the system (a mere $0.9 \, \Gyr$) requires $\sbp \sim 10^{-4}$. A similarly high degree of dissipation is required in the star, if stellar tides are be held responsible.

\begin{figure}
    \includegraphics[width=\columnwidth]{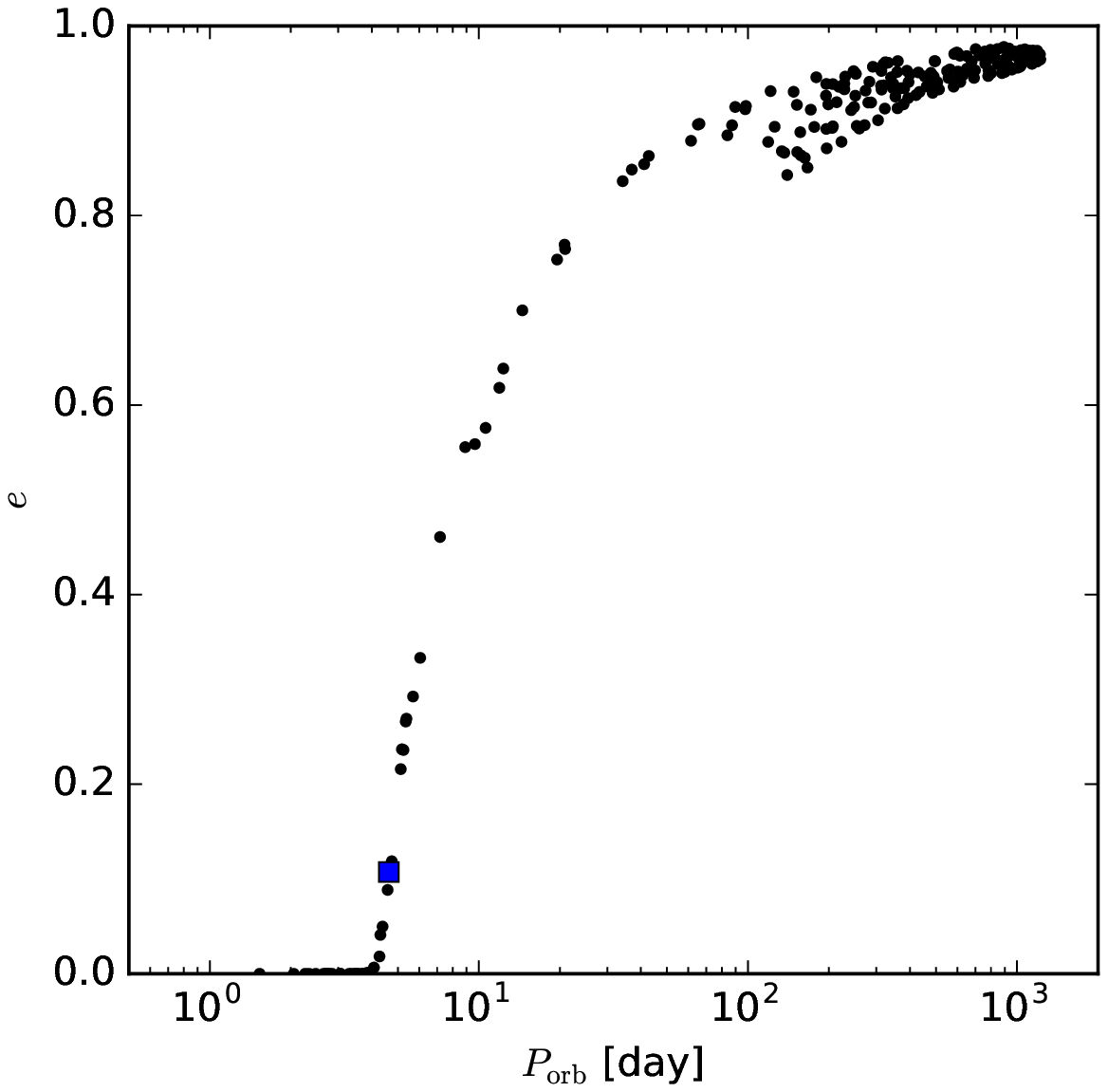}
    \caption{A period--eccentricity diagram showing the current orbit of HAT-P-14b (blue square) and a population of 250 model planets. The initial distribution of the planets is meant to represent plausible progenitors of HJs formed by high-eccentricity migration: all planets had initial periastron distances $a (1 - e) \leq 0.1 \, \AU$ and initial apastron distances $1 \, \AU \leq a (1 + e) \leq 5 \, \AU$.}
    \label{fig:HEM_HATP14}
\end{figure}

The discrepancy in our calibration for this system seems to be robust to the initial distribution of model planets. When we repeat our calibration procedure for a population designed to represent more realistically planets undergoing high-eccentricity migration (Figure \ref{fig:HEM_HATP14}), we derive a similar constraint on the dissipation constant,
$$ \sbp = (2 \pm 1) \times 10^{-4}. $$
A notable feature of Figure \ref{fig:HEM_HATP14} is that planets undergoing active circularisation are confined to a narrow region in the $P$--$e$ plane. This suggests an alternative means by which to constrain the dissipation constants, with accuracy comparable to the method we have chosen for the rest of this investigation.

Strictly speaking, this result is inconsistent with both constraints from H10 and the constraints yielded by other planets considered in this report. However, given the weaker rate of dissipation expected for more-massive planets ($\sbp$ being held constant) and the young age of the system, it may be that tidal dissipation simply has not begun in earnest for HAT-P-14b. Hence, the assumption that the planet lies on or near the $P$--$e$ envelope---upon which these calibrations are predicated---may be invalid. This uncertainty exemplifies the importance of finding more HJs around stars younger than $1$--$2 \, \Gyr$, so as to be able to discern the relative positions of HAT-P-14b and coeval systems in parameter space.

\subsubsection{WASP-18} \label{s:field:indiv:WASP18}

We now turn our attention to the more massive planets in the B17 sample. These planets have much more precisely measured eccentricities (sometimes with fractional errors of only a few percent), which helps to limit uncertainties. We will proceed with our analysis by initially assuming the same stellar tidal strength as before. H10 calibrated the stellar tide to this value by the same procedure we have undertaken now; this value matched the $P$--$e$ distribution of eccentric, massive planets known at the time, which was not especially sensitive to the planetary tide. This value also happens to agree very well with the value of $\sbs$ calculated by H12 for stars of WASP-18's mass and age.

WASP-18b \citep{hellier+09,southworth+09} is certainly the most extreme system we will consider, as it orbits in less than a day ($22.6$ hours) with a small, but nonzero, eccentricity of $0.0076 \pm 0.0010$ (B17). We find that planetary dissipation constants less than $\sbp = 4 \times 10^{-6}$ are consistent with the orbit of WASP-18b at the minimal system age of $0.5 \, \Gyr$. If the system is closer to $1.5 \, \Gyr$ old, then the constraint from below is around $\sbp = 6 \times 10^{-7}$. This is encouraging in that it is consistent with the ensemble models for planets in both mass bins, suggesting that WASP-18b's nonzero eccentricity is due more to the system's relative youth than to an incompatibility with the dissipation constants.

Assuming a planetary dissipation constant of $\sbp = 1 \times 10^{-6}$ (the geometric mean of the upper and lower limits on this quantity), we investigate the stellar dissipation in the same way, obtaining limits of
$$ 1.0 \times 10^{-7} < \sbs < 7.2 \times 10^{-7}. $$

\subsubsection{WASP-89} \label{s:field:indiv:WASP89}

The $6 \MJ$ planet WASP-89b \citep{hellier+15} has an eccentricity of $e = 0.193 \pm 0.010$ (B17) on an orbit some three days long. It is of interest to us because its host is a K-type star, rather than an F-type like WASP-18. As K-type stars have deep convective envelopes, this system may yield a different constraint on the stellar tide. Adopting a compromise value of $\sbp \sim 8 \times 10^{-7}$ between the constraints from CoRoT-16b and WASP-18b, we scan values of $\sbs$ instead, finding
$$ 7 \times 10^{-6} < \sbs < 2 \times 10^{-5}. $$
This is indeed substantially stronger than the dissipation inferred in F-type stars.

When the stellar dissipation is held constant at $\sbs = 1 \times 10^{-5}$, the planetary dissipation is constrained to be
$$ 5.0 \times 10^{-7} < \sbp < 5.0 \times 10^{-6}, $$
which is consistent with the other planets in our sample.

\subsection{Synthesis}

Taken together, the orbits of the transiting giant planets examined by B17 can be consistently described by planetary dissipation constants
$$ 6 \times 10^{-7} < \sbp < 1 \times 10^{-6}. $$
Values less than this allow for giant planets to have orbits at least as eccentric as those in the transiting sample, as well as for more-eccentric planets that may not have been discovered yet. In the main body of the text, we have adopted $\sbp = 8 \times 10^{-7}$ for dissipation in a giant planet.

\end{document}